\documentclass{emulateapj}
\usepackage{natbib}
\usepackage{times}
\usepackage{graphicx}
\usepackage{amsmath}
\usepackage{lscape}
\newcommand{\oii}{[{\rm OII}]\lambda 3727}
\newcommand{\oiiia}{[{\rm OIII}]\lambda 5007}
\newcommand{\oiiib}{[{\rm OIII}]\lambda 4959}
\newcommand{\oiii}{[{\rm OIII}]\lambda \lambda 4959,5007}
\newcommand{\neiii}{[{\rm NeIII}]\lambda 3869}
\newcommand{\niiha}{\log([{\rm NII}]\lambda 6583/H\alpha)}

\begin{document}

\title{The WHIQII Survey: Metallicities and Spectroscopic Properties of Luminous Compact Blue Galaxies}
\shorttitle{WHIQII Survey}

\keywords{Galaxies: observation --- Galaxies: starburst --- Galaxies: intermediate redshift}
\author{Erik J. Tollerud\altaffilmark{1}, Elizabeth J. Barton\altaffilmark{1}, Liese van Zee\altaffilmark{2},Jeff Cooke\altaffilmark{1,3}}
\altaffiltext{1}{Center for Cosmology, Department of Physics and Astronomy, The University of California at Irvine, Irvine, CA, 92697, USA}
\altaffiltext{2}{Astronomy Department, Indiana University, 727 East 3rd Street, Bloomington, IN 47405, USA}
\altaffiltext{3}{California Institute of Technology, 1200 East California Boulevard, Pasadena, CA 91125, USA}

\begin{abstract}

As part of the WIYN High Image Quality Indiana Irvine (WHIQII) survey,
we present 123 spectra of faint emission-line galaxies, selected to focus on intermediate
redshift ($.4 \lesssim z \lesssim  .8$) galaxies with blue colors that appear
physically compact on the sky. The sample includes 15 true Luminous
Compact Blue Galaxies (LCBGs) and an additional 27 slightly less
extreme emission-line systems.  These galaxies represent a highly
evolving class that may play an important role in the decline of star
formation since $z \sim 1$, but their exact nature and evolutionary
pathways remain a mystery.  Here, we use emission lines to determine
metallicities and ionization parameters, constraining their
intrinsic properties and state of star formation. Some LCBG
metallicities are consistent with a ``bursting dwarf'' scenario, while
a substantial fraction of others are not, further confirming that
LCBGs are a highly heterogeneous population but are broadly 
consistent with the intermediate redshift field.
In agreement with previous studies, we observe overall evolution in the
luminosity-metallicity relation at intermediate redshift.  Our
sample, and particularly the LCBGs, occupy a
region in the empirical R$_{23}$-O$_{32}$ plane that differs from
luminous local galaxies and is more consistent with dwarf Irregulars
at the present epoch, suggesting that cosmic ``downsizing'' is
observable in even the most fundamental parameters that describe star
formation.  These properties for our sample are also generally consistent 
with lying between local galaxies and those at high redshift, as expected by
this scenario.  Surprisingly, our sample exhibits no detectable correlation
between compactness and metallicity, strongly suggesting that at these
epochs of rapid star formation, the morphology of compact star-forming
galaxies is largely transient.

\end{abstract}

\section{Introduction}
\label{sec:intro}
Faint blue galaxies have long been a subject of study \citep[e.g.,][and references therein]{koo94,ellis97}. The most extreme class, Compact Narrow Emission Line Galaxies (CNELGs) were first noted by \citet{koo94}. Generally speaking, they are luminous ($\lesssim B_*$) and starbursting as indicated by blue optical colors (i.e. B-V $\lesssim 0.6$).  They have bright, narrow emission lines, are compact (i.e. $r_h \lesssim 3$ kpc), and are predominantly found at intermediate redshift ($0.4 < z < 1.2$). Later studies of these objects \citep[e.g.,][]{guz98,bvz06cnelg} and similar populations such as luminous compact galaxies \citep[e.g.,][]{ham01} and Luminous Compact Blue Galaxies \citep[LCBGs, e.g.,][]{noeske06}, have revealed that these galaxies have very high star formation rates, perhaps -- depending on the definitions used -- representing up to $\sim 45\%$ of total star formation at intermediate redshift, and 20\% of the field number density \citep{phillips97,guz97,noeske06}.  At the present epoch, however, these galaxies are far less common \citep{koo94,phillips97,guz97,werk04,noeske06}, suggesting that these objects may play an important role in the global decline of star formation \citep{madau96} and the evolution in the blue luminosity function \citep{lilly95,ellis97}.  Despite this, it remains unknown what their end state is in the local universe.  While there are subtle differences in the terminology used in the literature, we follow the definition of \citet{noeske06}, and use the ``LCBG'' term, as essentially all intermediate redshift objects meeting LCBG criteria also show narrow emission lines.

Two primary scenarios have been suggested for the evolutionary scenario of LCBGs.  The ``bursting dwarf'' hypothesis \citep[e.g.,][]{koo95,guz98,hoyos05} suggests that LCBGs are low mass galaxies undergoing a single major starburst that will exhaust their gas.  They then fade after the starburst by several magnitudes into present day dwarf spheroidal or elliptical (dSph/dE) galaxies.  While it is possible that underlying pre-starburst stellar populations might exist that prevent the galaxies from fading this much, the scenario still predicts low metallicities, consistent with dwarf galaxies rather than bright galaxies.  The alternative interpretation is that the LCBGs are {\it in situ} bulge-formation of lower luminosity spirals \citep[e.g.,][]{ham01,bvz06cnelg}.  The bulges-in-formation scenario predicts higher metallicities than those likely to be observed in local dwarf galaxies \citep{KZ99}. The bulge and dwarf scenarios are not mutually exclusive -- the LCBGs may be a heterogeneous population, or, i.e. evolve to become mostly lower luminosity spirals now \citep{bvz01,ham01,noeske06,bvz06cnelg}.

The WIYN High Image Quality Indiana Irvine (WHIQII) survey aims to address the question of the evolutionary paths of LCBGs by examining the metallicities of a large sample derived using the strong $\oii$, $\oiiia$, $\oiiib$, and $H\beta$ lines; primarily through the use of the $R_{23} = (\oii+\oiii)/ H\beta$ and $O_{32} = \oiii / \oii$ indicators (see \S \ref{sec:met}).     This paper is organized as follows: in \S \ref{sec:isamp} we describe the imaging and initial sample selection.  In \S \ref{sec:dat}, we discuss observation and reductions of of the spectroscopic survey and selection of the sample used for analysis.  In \S \ref{sec:numden} we briefly address the number density of LCBGs determined from WHIQII. In \S \ref{sec:met} we discuss the metallicities of the sample and how they are derived as well as the luminosity-metallicity (LZ) relation.   \S \ref{sec:disc} discusses trends and local comparisons, and in \S \ref{sec:conc} we present our conclusions.  Where relevant, we assume a WMAP5 \citep{kom09WMAP} $\Lambda$CDM cosmology.

\section{Imaging and Sample Selection}
\label{sec:isamp}
The BVRI imaging data for WHIQII was obtained from observations using the Mini-Mosaic Imager (MiniMo) on the WIYN 3.5m telescope over 25 nights from 2001-2004 under photometric conditions in excellent (median $\sim 0.7"$) seeing.  The details of the photometric reductions will be described in a forthcoming paper (van Zee et al. 2009, in prep), although we describe our determination of the half-light radius ($r_h$) for this paper in section \S \ref{sec:samp}.  

Our photometric sample selection was guided by a focus on finding LCBGs and
related systems at intermediate redshift.  Using the slitmask
design software AUTOSLIT\footnote{http://www2.keck.hawaii.edu/inst/lris/autoslit.html}, we targeted objects with R-band magnitudes in the
range $21 < {\rm m_R} < 23$ that had reliable photometric
redshifts of $z < 1$.  The first priority bin focused on small,
(observed-frame) blue objects with full-width half maxima (FWHM) $\leq
0.^{\prime\prime}8$ and $V-I < 1.6$.  The second priority bin relaxed
the size requirements to ${\rm FWHM} \leq1.^{\prime\prime}5$ but kept the blue color
requirement.  For the third priority bin, we relaxed the color
requirement to $V-I < 2$ and the fourth priority bin allowed objects
with $1.^{\prime\prime}5 < {\rm FWHM } < 2^{\prime\prime}$.  Finally,
we filled the remaining mask space with galaxies of any size and colors
$V-I < 2$. From the 11 $9.6^\prime \times 9.6^\prime$ fields imaged, this procedure resulted in 
213 objects that were selected for spectroscopic follow-up.

The data  in the WHIQII survey falls into three samples:
the entire sample of galaxies photometrically identified in the
WHIQII imaging survey, referred to as the ``photometric'' hereafter,
the objects in the spectroscopic survey for which a redshift can be
determined (the ``spectroscopic'' sample), and the portion
of the spectroscopic survey that also meet the LCBG criteria (the
``WHIQII LCBG'' sample).  Probable AGN are removed from the 
latter two samples as described in \S \ref{sec:samp}.

\section{Spectroscopic Data}
\label{sec:dat}

\subsection{Observations and Reductions}
\begin{figure}[htbp!]
\plotone{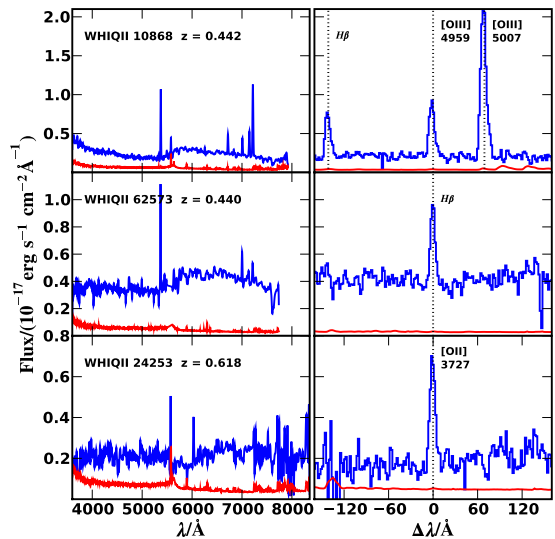}
\caption{Three sample spectra of LCBGs from the WHIQII spectroscopic sample.  Upper spectrum (blue) is the science spectrum, while the lower (red) is the per-pixel standard deviation. Left column is the whole spectrum smoothed by an 8 \AA \: Gaussian filter, while the right column zooms in on selected spectral features with the x-axis indicating the  separation from that feature in \AA.}
\label{fig:sampspec}
\end{figure}

Spectroscopic observations of 213 targets were taken on the nights of February 11, 2005  and August 3-4, 2005 with the dual-channel Low Resolution Imaging Spectrometer \citep[LRIS,][]{oke95lris,mccarthy98lrisB} on the Keck I telescope under moderate conditions.  The red side grating was 400 line mm$^{-1}$ blazed at  8500 \AA, with a blue side grism of 400 line mm$^{-1}$ blazed at 3400 \AA, along with the 5600 \AA \: dichroic.  We reduce the spectra using standard IRAF\footnote{IRAF is distributed by the National Optical Astronomy Observatories, which are operated by the Association of Universities for Research in Astronomy, Inc., under cooperative agreement with the National Science Foundation.} tasks in the PyRAF\footnote{PyRAF is a product of the Space Telescope Science Institute, which is operated by AURA for NASA} environment.  We bias subtract and flat field each spectrum.  Twilight flats are also applied to the spectra to correct for uneven slit illumination, which also serves to remove small amounts of scattered light.  We determine a wavelength solution by identifying lines from the  HeNeAr (red side) and ZnCd (blue) arc spectra.  Care must be taken in choosing the extraction aperture for faint emission line objects, because the goal is to maximize signal-to-noise in emission lines while minimizing the absorption spectrum from the underlying stellar continuum.  We accomplish this choice by identifying objects with both strong emission lines and continua, to determine the optimal aperture width.  The resulting curves of growth are approximately the same for all objects, so we select a fixed width.  This procedure also simplifies matching the red and blue spectra by allowing for a fixed aperture size ratio as determined by the plate scale.    Finally, we flux calibrate the science spectra with spectrophotometric standard stars (Feige 110, Feige 34, BD284211, G191B2B).   Figure \ref{fig:sampspec} shows example WHIQII spectra.  We note that while matching of the blue and red arms can be problematic, nearly all of our spectra with wavelength overlap match continuum luminosities on the two arms, as evidenced by mismatches lower than the RMS noise floor (see Figure \ref{fig:sampspec}).  Those few for which there is a substantially mismatch generally have very low continuum SNR, so the mismatch in those cases is unsurprising.

\begin{figure}[htbp!]
\plotone{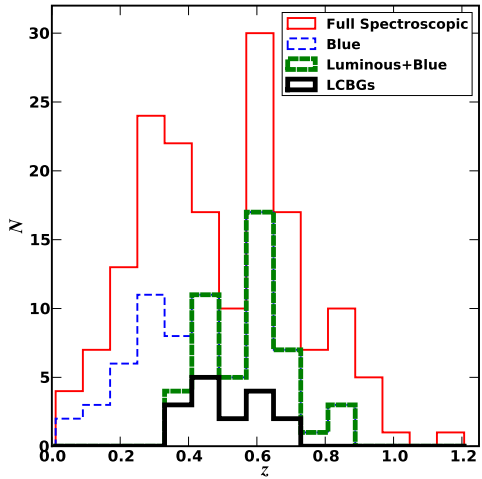}
\caption{Redshift distribution of WHIQII spectroscopic sample. Thin solid (red) lines are for the entire sample of 168 objects with identifiable spectral features, thick solid (black) are objects that fit the selection criteria of LCBGs, thick dashed (green) are those that meet the luminosity and compactness criteria (but are not necessarily blue), and thin dashed (blue) are objects that make the color cut and are compact but are not necessarily luminous enough (see \S \ref{sec:samp}).}
\label{fig:redshift}
\end{figure}

We visually examine the reduced spectra for spectral features and measure fluxes as well as equivalent widths for each feature using a custom software package spylot\footnote{http://ps.uci.edu/$\sim$etolleru/software.html\#spylot}.  Of the 213 spectra taken, 168 show strong enough spectral features to be assigned a redshift, and 123 have unambiguous emission lines. Figure \ref{fig:redshift} presents the redshift distribution of these objects.  For those with secure redshifts, we determine K-corrections using the  Kcorrect v4.1.4 \citep{b07kcorr} code on the photometry assuming the spectroscopically determined redshifts.  We correct for Galactic extinction and reddening of the spectral features using the dust maps of \citet{sfd98} along with a \citet{cardelli89} extinction curve.

Because H$\alpha$ cannot be measured in the rest-frame optical for intermediate redshift objects, and H$\gamma$ is undetected for essentially the entire sample, there is no reliable way to correct emission line fluxes for internal reddening.  Hence, we primarily use equivalent width ratios to measure metallicity indicators instead of the direct flux ratios to attempt to correct for this effect \citep[see ][ for more details on this approach]{KK04}.  In the cases where H$\beta$ emission is seen superimposed on stellar absorption, a Gaussian model for the absorption is fit, with the emission line flux measured above this fit.  It is likely that in some cases, this absorption is present but undetected or poorly resolved, and hence the actual emission of H$\beta$ is likely systematically higher than the measured flux.  The typical magnitude of this effect is not enough to significantly affect our key results, but it is a source of systematic uncertainty.   As a final general note, wherever possible, we analyze the comparison sample data described below in the same way as the WHIQII data to minimize induced scatter from many layers of corrections.

\subsection{AGN Contamination and Sub-Samples}
\label{sec:samp}

\begin{figure}[htbp!]
\plotone{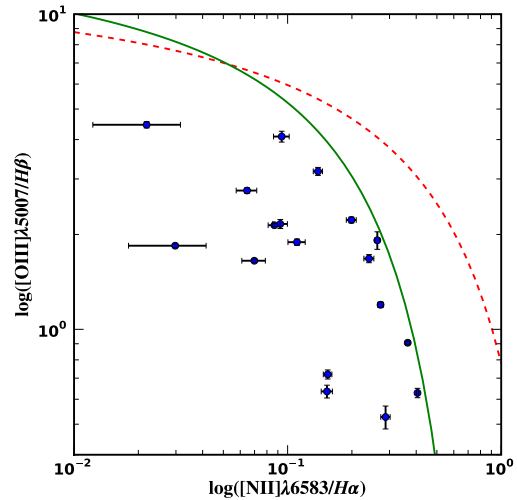}
\caption{BPT diagram for WHIQII objects with detectable $H\alpha$ and $[N{\rm II}]$ emission (blue circles with error bars).  Note that this sample is biased to the low redshift end of WHIQII due to the inaccessability of H$\alpha$ and [NII]$\lambda 6583$ at higher redshift. The solid line (green) corresponds to the empirical AGN/starburst line of \citet{kauf03}, while the dashed line (red) is the ``maximal starburst'' model of \citet{kewley01}.}
\label{fig:BPT}
\end{figure}

It is important to address is the possibility of sample contamination by Active Galactic Nuclei (AGN), as AGN show emission line spectra that can mimic starbursts but do not reflect the properties present in the star forming regions.  Because the selection criteria emphasize luminous, compact objects, there is potential for a serious contamination issue.  We address the question of contamination with the BPT diagram \citep{BPT81}; we plot $\niiha$ vs. $\log([{\rm OIII}]\lambda 5007/H\beta)$. Use of this diagram is complicated by the fact that most of the WHIQII spectra are at a redshift such that $H\alpha$ and $[N{\rm II}]$ lie within the OH forest or are outside the wavelength range covered, and hence cannot be placed on the diagram.  Figure \ref{fig:BPT} plots the few WHIQII objects for which all these lines are measurable, along with the empirical delineation of \citet{kauf03} and the ``maximal starburst'' model of \citet{kewley01}. The WHIQII galaxies are clearly all on the star-forming side of this diagram, indicating that at least the lower redshift side of the sample has little or no AGN contamination.  

\begin{figure*}[htbp!]
\plottwo{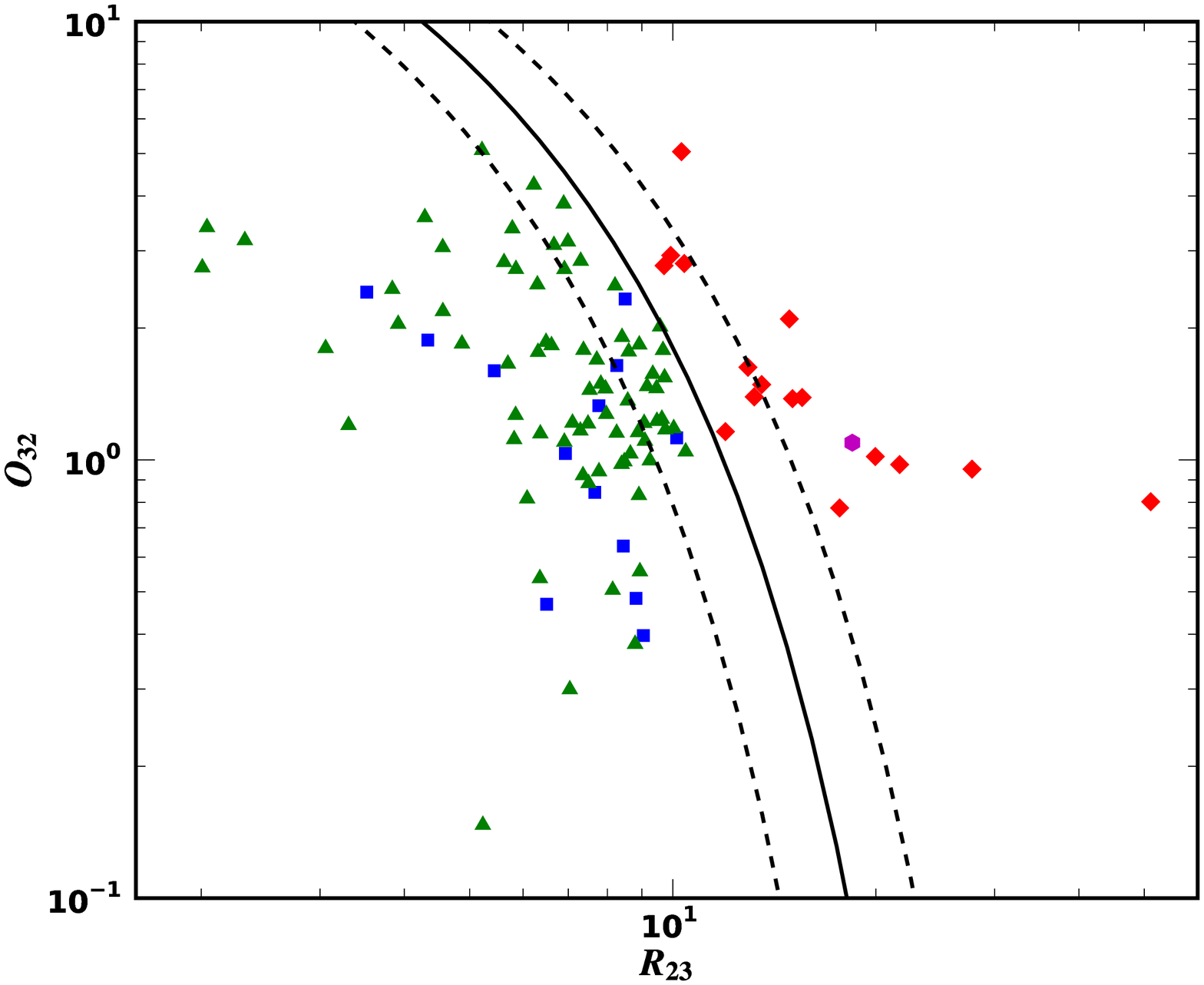}{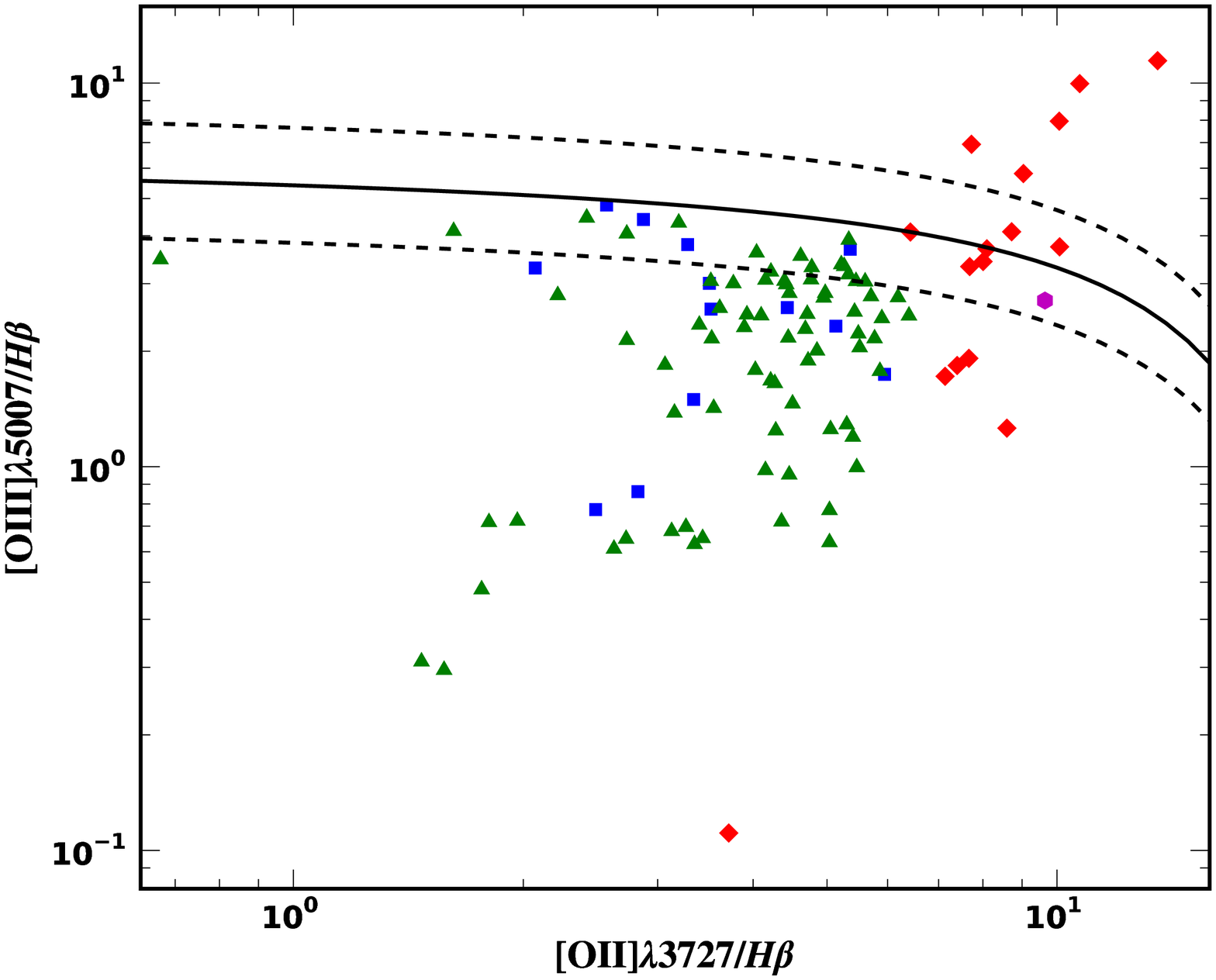}
\caption{Diagnostic diagrams from \citet{lamar04} for AGN contamination using strong lines detectable at intermediate redshift.  In both diagrams, objects to the lower-left of the solid line are star-forming galaxies, while object to upper-right are AGN, with transition zones delineated by the dashed lines.  Squares (blue) are WHIQII objects that meat the criteria to be LCBGs, while triangles (red) are objects that do not.  Hexagons (magenta) are objects removed from the LCBG catalog, while diamonds (red) are those removed from the full WHIQII catalog.} 
\label{fig:lamar}
\end{figure*}

\citet{lamar04} provide two diagnostic techniques better suited to intermediate redshift by making use of only strong lines at shorter wavelengths.  The associated diagrams for WHIQII are shown in figure \ref{fig:lamar}.  These diagrams show that while most objects in the spectroscopic sample lie within the star-forming region of the diagram, 19 lie to the right of the demarcation line and hence are more consistent with AGN.  Hereafter, when referring to the spectroscopic or LCBG sample, we first remove objects that appear to be AGN based on these criteria before all other analysis.  However, we note that a significant fraction (36 objects, or $35 \%$ of the objects after removing the probable AGNs) are still within the range of the scatter observed by \citet{lamar04} and hence some fraction may be low-luminosity AGNs mixed with a starburst.

Having cleaned the sample of probable AGN we now move onto to selection of LCBGs in the spectroscopic sample, given the information from the spectra.  There are various definitions of LCBGs in the literature, but for our purposes we follow \citet{noeske06} requirements that $B-V < 0.6$, $B < -18.5$, and $r_h<3.5$ kpc.  While the first two are basic photometric parameters, the question of determining compactness is less straightforward.  

To determine $r_h$, we use the FWHM measured in SExtractor \citep{SExtractor} assuming a Gaussian profile for the photometric object in the R band ($w_{obs}$), subtract the seeing in quadrature, and correct this to the radius corresponding to half the flux of a Gaussian profile.  Correcting this angular size with the angular diameter distance ($d_a$) for each galaxy gives an approximation to the physical $r_h$.  Hence, for each galaxy, we have 
\[
r_h =  \frac{d_a}{2}\frac{\sqrt{w_{obs}^2-w_{seeing}^2}}{206265"} {\rm kpc}.
\]

\begin{figure*}[htbp!]
\plotone{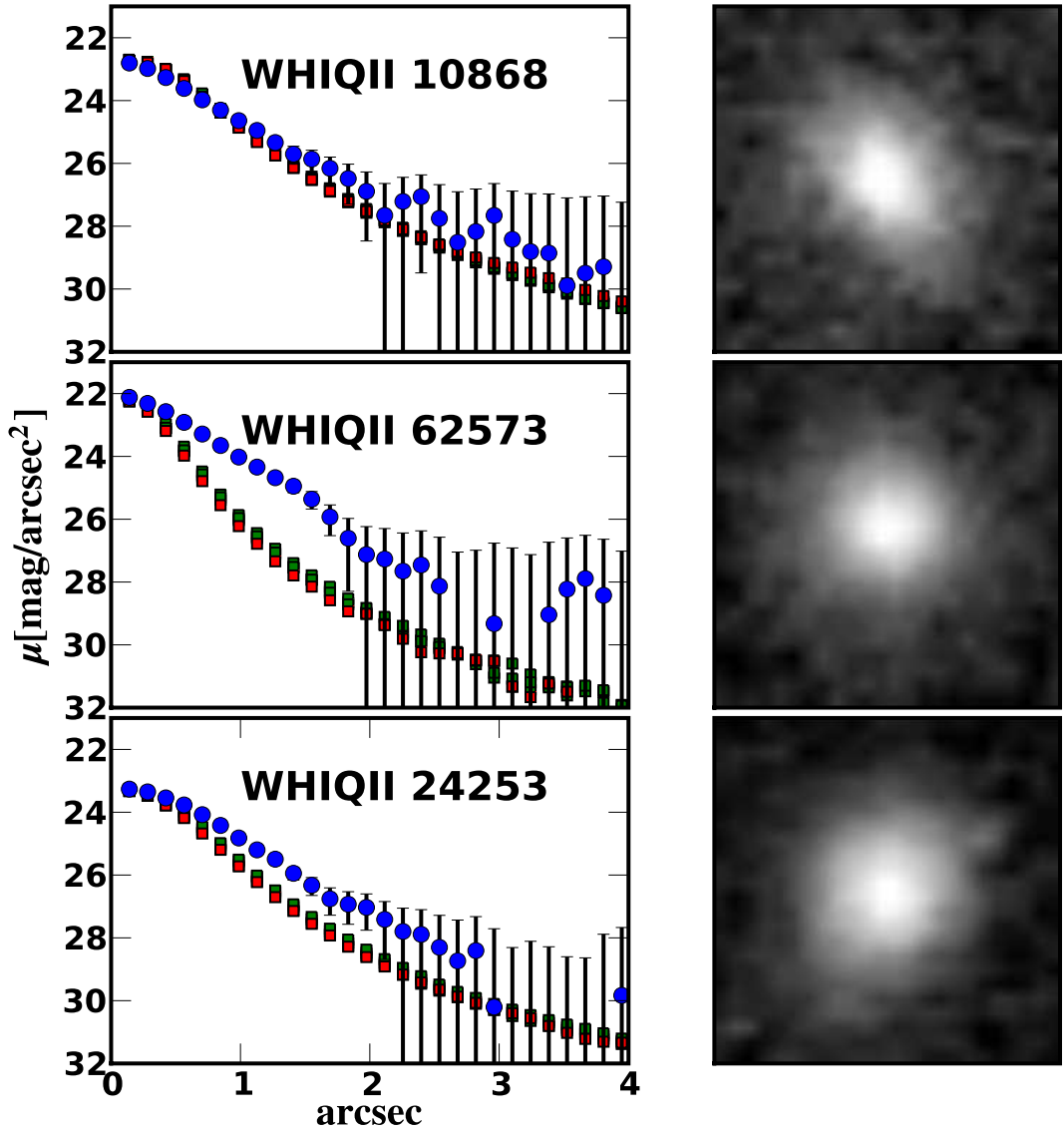}
\caption{Surface brightness profiles (left) and images (right) of selected WHIQII objects.  From top to bottom are WHIQII 10868, WHIQII 62573, and WHIQII 25253. In surface brightness profiles, (blue) circles with error bars are the object, and squares are PSF stars (red = nearest, green = others)}
\label{fig:ims} 
\end{figure*}

\begin{figure}[htbp!]
\plotone{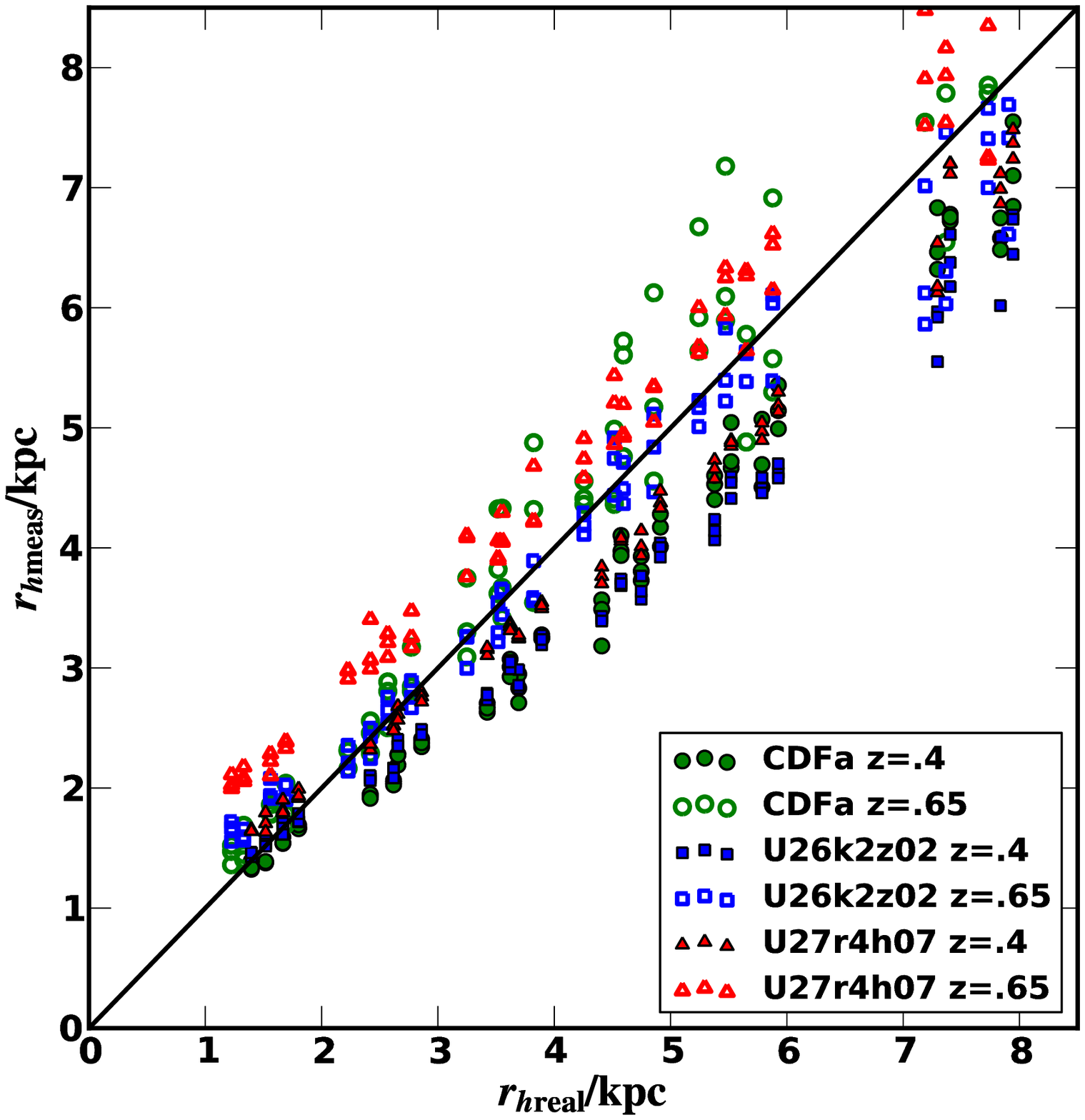}
\caption{Model half-light radius vs. measured half-light radius (following the procedure described in the text) for model galaxies added to WHIQII fields.  Green circles are for models placed in the CDFa field (median seeing of $.56\arcsec$ FWHM), blue squares are U26k2z02 (median $.73\arcsec$), while red traingles are U27r4h07 ($1.0\arcsec$).  This spans the range of seeing conditions encountered in the survey.  Solid points are those for which the model galaxy was assumed to be at a redshift of 0.4, typical for the survey, while open circles are redshift 0.65, the upper range for WHIQII LCBGs.  The black line is the one-to-one relation. }
\label{fig:modelsims}
\end{figure}

This procedure is somewhat simplistic, although motivated by the compactness of LCBGs that is apparent from the images and surface brightness profiles presented in Figure \ref{fig:ims}.  Hence we performed simulations to test the validity of this method by generating model galaxies composed of a spherical bulge component \citep[with a ][ profile]{DeVaucouleurs} and an exponential disk at a variety of assumed inclination angles.  We generate a grid of models with varying bulge-to-disk ratios, redshifts, and half-light radii, add noise matching the CCD characteristics, convolve with the seeing appropriate for the field, and add the resulting model onto a blank area of a WHIQII field.  We then re-extract sources from the field and follow the procedure described above on the new object to determine an estimate for $r_h$. Figure \ref{fig:modelsims} compares the resulting half-light radius to the actual model half-light radius for three fields that span the range of seeing conditions in the survey.  

While the points scatter about the one-to-one relation, most of the scatter is due to a combination of variation of properties with redshift (higher redshift objects come closer to the PSF and hence appear more concentrated than they should be) and inclination angle (highly inclined disks tend to be higher surface brightness and hence puff up the Gaussian fit).  The other properties varied across the simulation have much weaker effects, although they contribute to the overall scatter.  The highly inclied objects, when visually inspected, however, appear very different from LCBGs.  For objects that meet the luminosity criterion and are within the redshift range of LCBGs, the large disk would be clearly visible in the images, which is not the case for all the LCBGs in the WHIQII sample.  

\begin{figure}[htbp!]
\plotone{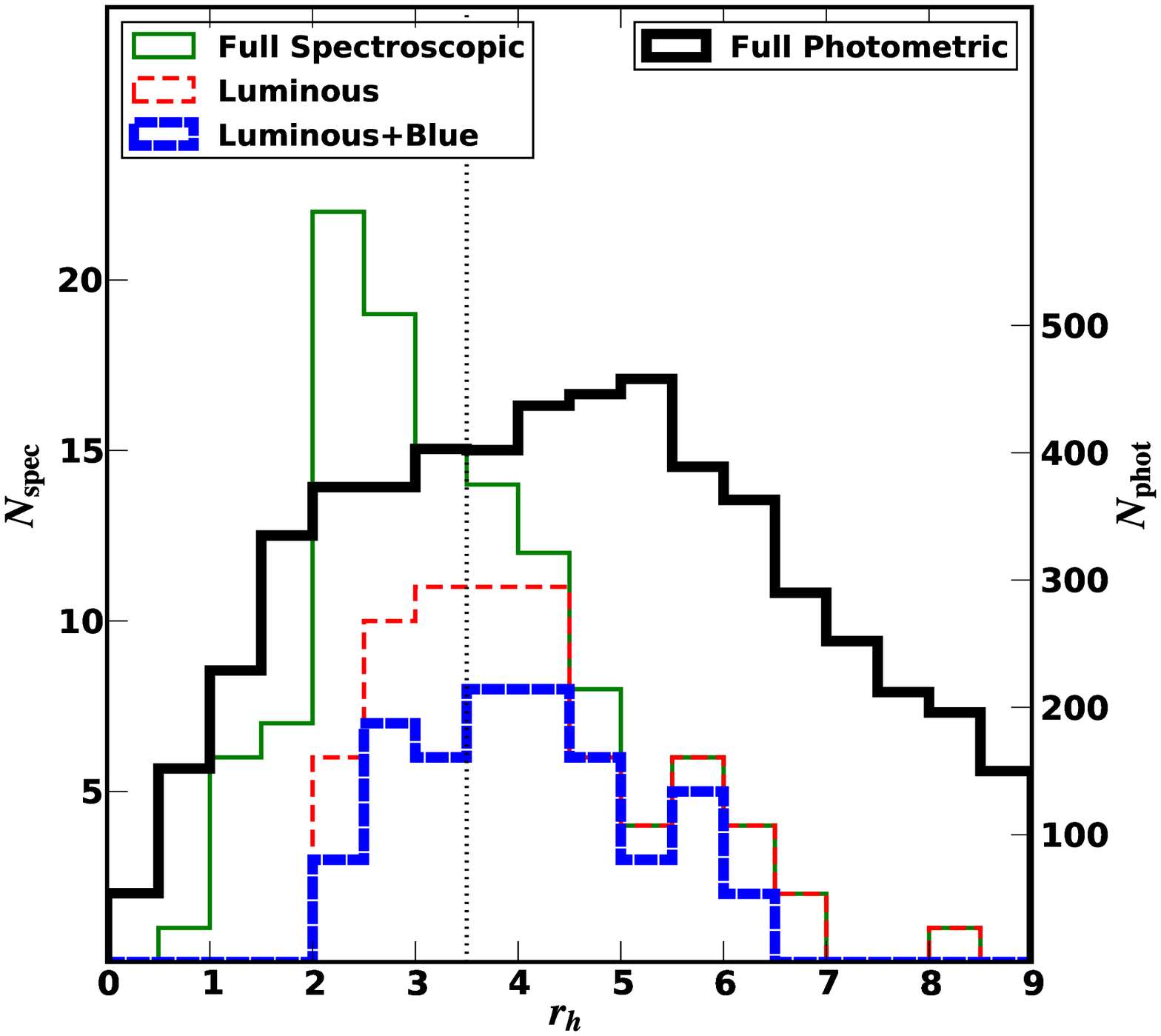}
\caption{Half-light radius distribution for WHIQII objects.  The thick solid (black) line is $r_h$ from the portion of the WHIQII photometric sample with photometric redshifts in the range $0.4 < z < 1$ corrected with angular diameter distances derived from the photometric redshift.  The this solid (green) line represents the the WHIQII spectroscopic sample, the thin dotted (red) line are the sub-sample of those that meet the luminosity cut ($B<-18.5$), and the thick dotted (blue) line represents the sub-sample that is both luminous and blue ($B-V<.6$). The vertical dotted line corresponds to 3.5 kpc, the compactness criterion we adopt for LCBGs.  Hence, LCBGs are objects from the dashed (blue) sample that are left of the vertical line.  The right y-axis is for the photometric sample (solid/black line), while the left y-axis is for the spectroscopic (all others). }
\label{fig:rhdist}
\end{figure}

A we consider a realistic sample of LCBGs consists of the lower inclination objects and/or the objects with low bulge-to-disk ratios.  In this population there is a slight ($\sim 10 \%$) systematic bias for the measured $r_h$ to lie lower than the actual $r_h$.  However, the only range in which our absolute value of $r_h$ matters is for the LCBG/non-LCBG categorization.  At the boundary ($r_h = 3.5$ kpc) the offset is within the scatter.  Thus, the bias is not a concern for the trends we are interested in searching for.  We quantify the scatter by measuring the residuals from a linear fit to the lower inclination and lower B/D objects and find a standard deviation of$.43$ kpc for the range of $r_h$ shown in Figure \ref{fig:rhdist}.  We adopt this as our error in $r_h$ for analysis in \S \ref{sec:disc}.

The distribution of $r_h$ for actual WHIQII objects is shown in Figure \ref{fig:rhdist} for both the WHIQII photometric sample and the WHIQII spectroscopic sample.  For the former, distances are determined from photometric redshifts derived via standard techniques that will be described in a forthcoming paper, and for the latter, spectroscopically determined redshifts are used. 

With the definition outlined above, our sample thus contains 15 LCBGs, and about as many near the boundaries of that categorization.  Figure \ref{fig:redshift} shows that these criteria already force the sample to lie within the intermediate redshift range ($0.4 \lesssim z \lesssim 1$), validating the criteria we use to select intermediate redshift LCBGs from the photometric sample.  As the distributions in Figures \ref{fig:redshift} and \ref{fig:rhdist} show, there are a large number of objects in the WHIQII survey that have clear redshifts, but still do not meet the LCBG criteria.  Some of these show very weak or no emission (redshifts are from absorption features),  and there are 89 objects that have strong emission lines but are not AGN.  Of these, 50 have $z<0.4$, and most are dwarf galaxies that are intrinsically compact but are not LCBGs because of low luminosity.  The LCBG luminosity cut essentially guarantees that the objects are at intermediate redshift, although 24 of the objects meeting the luminosity cut are not blue enough, leaving 42 objects that are luminous and blue.   While only 15 are LCBGs as defined, the rest of the objects (27) are near the compactness threshold -- thus, the entire WHIQII sample is biased to galaxies much more compact than typical galaxies, even if they do not necessarily meet the official criterion described above.  Hence, we often will also present objects that do not meet the compactness criterion, but do meet the color and luminosity criteria, as that sample still represents near-LCBGs. It is also important to note that there may be some extremely compact LCBGs that we have missed that are smaller than the PSF, but the fact that most of the objects near the PSF size are below or near the lower luminosity limit in the redshift range we are sensitive to suggests that this is probably a small fraction.

Tables \ref{tab:photdata} and \ref{tab:specdata} give a summary of all objects in the WHIQII spectroscopic sample, along with their relevant properties. Note that in this table we make use of the \citet{bdj01mstar} relations to compute $M_*$.  This paper assumes stellar populations for normal spirals and hence are not fully correct for the starburst populations likely to be present in LCBGs.  However, the values should still be a rough estimate of the stellar mass so as to remove the first order bias for starbursts appearing overly luminous in bluer bands.

\section{LCBG Fraction}
\label{sec:numden}
To get an estimate for the number density of LCBGs in this redshift range, we first construct the overall galaxy population in the WHIQII photometric sample in the $0.4 < z < 1$ redshift range. We filter stars using the S/G parameter in SExtractor \citep{SExtractor}.  We determine the cutoff value ($S/G \leq 0.98$) by inspecting the S/G values for the spectroscopically identified emission line galaxies and then visually examine the objects that passed and failed this criterion to ensure that they are primarily extended objects or PSF-like on opposite sides of the cutoff.  We produce K-corrected magnitudes for the remainder of the photometric sample using the photometric redshifts, filtering out a small number of values for which the K-corrected magnitudes are highly discrepant.  This eliminates poorly deblended objects or those with artifacts as well as spurious features such as noise and cosmic rays that were missed in the earlier steps of the photometric reduction.  We then consider only the objects for which $0.4 < z_{\rm phot} < 1$, producing a cleaned intermediate redshift galaxy sample ($N_{\rm phot} = 1744$).  Next, we determine the completeness limit as the apparent magnitude (in the most sensitive band, R) at which the histogram of apparent magnitudes begins to roll over.  At $z = 1$, this corresponds to a limiting absolute magnitude of $R=-18.5$.  This is close to the luminosity cut for LCBGs, so we apply this limit ($B < -18.5$) to the K-corrected absolute magnitudes to generate a volume-limited sample. To these, we apply the LCBG criterion: $r_h < 3.5$ and $B-V < 0.6$, producing a photometric LCBG sample $N_{\rm LCBG}=199$.  Thus, we estimate an LCBG fraction of $f_{{\rm LCBG},B<-18.5} \approx 11\%$. Using the luminosity functions of \citet{faber07LFz1} for $z=0.5$ (near the spectroscopic WHIQII LCBG median redshift of .49), we can correct this fraction for fainter galaxies that are missed in WHIQII but that are definitively not LCBGs because of failing the luminosity criterion.  For a faint-end cutoff of $B=-17$,  $\frac{N(B<-18.5)}{N(-18.5<B<-17)} = 83 \%$, so the fraction of LCBGs for all galaxies brighter than $B<-17$ is $f_{{\rm LCBG},B<-17} \approx 10\%$.  For $B=-16$,  $\frac{N(B<-18.5)}{N(-18.5<B<-16)} = 43\%$, so $f_{{\rm LCBG},B<-16} \approx 5\%$.  These fractions are slightly lower than previous results \citep[e.g.,][]{guz97,noeske06}, likely resulting from a slightly lower median redshift combined with the fact that LCBGs clearly evolve very strongly at $z \lesssim 1$.  Nevertheless, these numbers show that LCBGs represent a significant fraction of all observable galaxies at intermediate redshift, despite being largely absent at the present epoch \citep{werk04}.

\section{Metallicity}
\label{sec:met}

The primary tracer of metallicity available from HII region emission lines is the oxygen abundance. The most direct means of measuring $\log(O/H)+12$ is via the electron temperature ($T_e$) method \citep[e.g.,][]{osterbrock, KZ99,yin07} from the $[{\rm OIII}] \lambda 4363$ emission line.  However, only a few galaxies in the WHIQII sample exhibit this line, as it is a weak transition, and they are low redshift ``contaminants'' with intrinsic luminosities too low to be considered LCBGs.   Most strong-line calibrations are unavailable for intermediate redshift objects, where the  $H\alpha$, $[NII]$, and $[SII]$ lines are overlapping on OH sky lines or outside the spectral range (see \citealt{KD08} for a recent review).  Thus, the primary estimator available is $R_{23}$  which we combine with $O_{32}$ as described below. In some cases, $\neiii$ is detected, allowing the use of the Ne3O2 estimator \citep{shi07}, although the scatter in this estimator is much higher than $R_{23}$ and does not appear to be consistent with the other indicators for our sample (see  \S \ref{sec:MZ}).  

\begin{figure}[htbp!]
\plotone{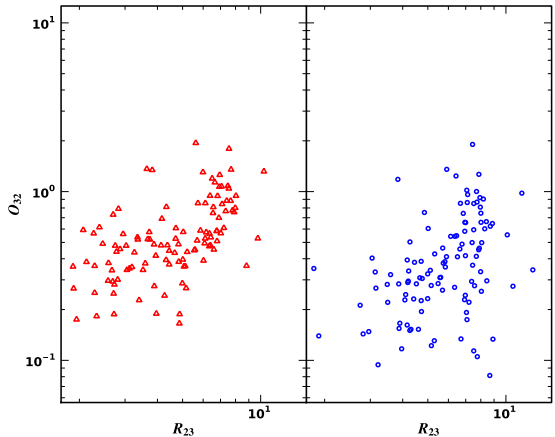}
\caption{$R_{23}$ vs. $O_{32}$ emission line flux ratios for objects from \citet{nfgs00}.  The left panel (red triangles) use uncorrected ratios, while in the right panel (blue circles), the fluxes in each line have been corrected for extinction via a \citet{calz94} extinction curve. }
\label{fig:extinctr23o32}
\end{figure}

\begin{figure}[htbp!]
\plotone{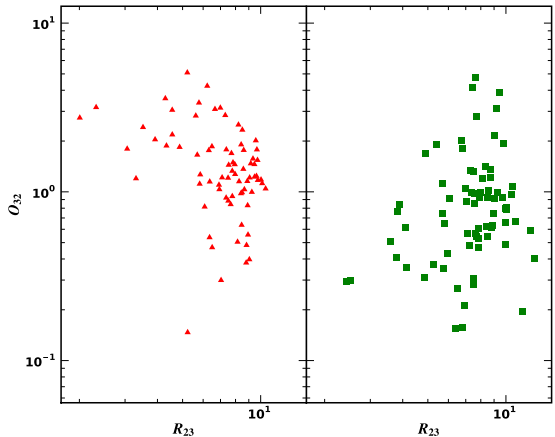}
\caption{$R_{23}$ vs. $O_{32}$ emission line ratios for objects with the necessary emission lines.  The left panel (red triangles) are line ratios from the WHIQII spectroscopic sample using the direct \emph{flux} ratios, while right (green squares) are ratios of \emph{equivalent width} measurements for the same set of objects.  }
\label{fig:r23o32fluxvsew}
\end{figure}

An important concern when using these estimators is internal extinction in the target galaxies.  Given that LCBGs exhibit massive amounts of star formation, it is quite plausible that the star formation is embedded in large amounts of dust and is thus prone to extinction and reddening that may alter the flux ratios. As an illustrative example, Figure \ref{fig:extinctr23o32} corrects the most important flux ratios ($R_{23}$ and $O_{32}$) from the local sample of \citet[][see \S \ref{sec:comp} for details on this sample]{nfgs00} for extinction using the Balmer decrement assuming a \citet{calz94} extinction curve.  Noting the clear offset in this plane and the lack of multiple Balmer lines to correct our sample, we follow \citet{KK04} and compare the direct measurement of the flux ratios to the ratios of the lines' equivalent widths in Figure \ref{fig:r23o32fluxvsew}.  It is immediately clear from this plot that the flux ratio measurements are discrepant from the equivalent width measurements in the same sense as the corrected flux ratios from Figure \ref{fig:extinctr23o32}.  Furthermore, comparison between Figures \ref{fig:r23o32fluxvsew} and \ref{fig:r23o32grids} show that the flux ratio points appear to be in parts of the metallicity grid that are improbable, as the implied metallicity would be offset by $\sim 1$ dex from expected values.  Hence, we conclude that extinction is quite significant for most of these galaxies, and use equivalent width ratios instead of direct flux ratios for the remainder of this paper (see \citealt{KK04} for validation of and more on using equivalent width for these line ratios).

\subsection{Metallicity and Ionization Parameter Calibrations}
The literature contains many oxygen abundance calibrations for the $R_{23} = (\oii+\oiii)/ H\beta$ estimator \citep{pagel79r23,MG91,Z94,KD02,KK04,pil05}. There are significant offsets (of order $\sim 0.3$ dex) between these calibrations, and even larger discrepancies between $R_{23}$ estimators and the $T_e$ method \citep{ken03,KK04,KD08}.  Changes in the adopted solar oxygen abundance \citep[e.g.,][]{asp04} can affect this discrepancy even further.  \citet{KD08} conclude that only \emph{relative} metallicities for a given calibration are reliable given the discrepancies. Hence, given the use of the \citet{KD02} calibration for a number of previous intermediate redshift studies, we use it as parametrized in \citet{KK04} for the remainder of this paper, taking care to use only that calibration when comparing to other results.  Note, however, that we repeat most of the analysis described below using the \citet{MG91} calibration as well as \citet{KD02}, and all of the qualitative results are unaffected.

\begin{figure*}[htbp!]
\plottwo{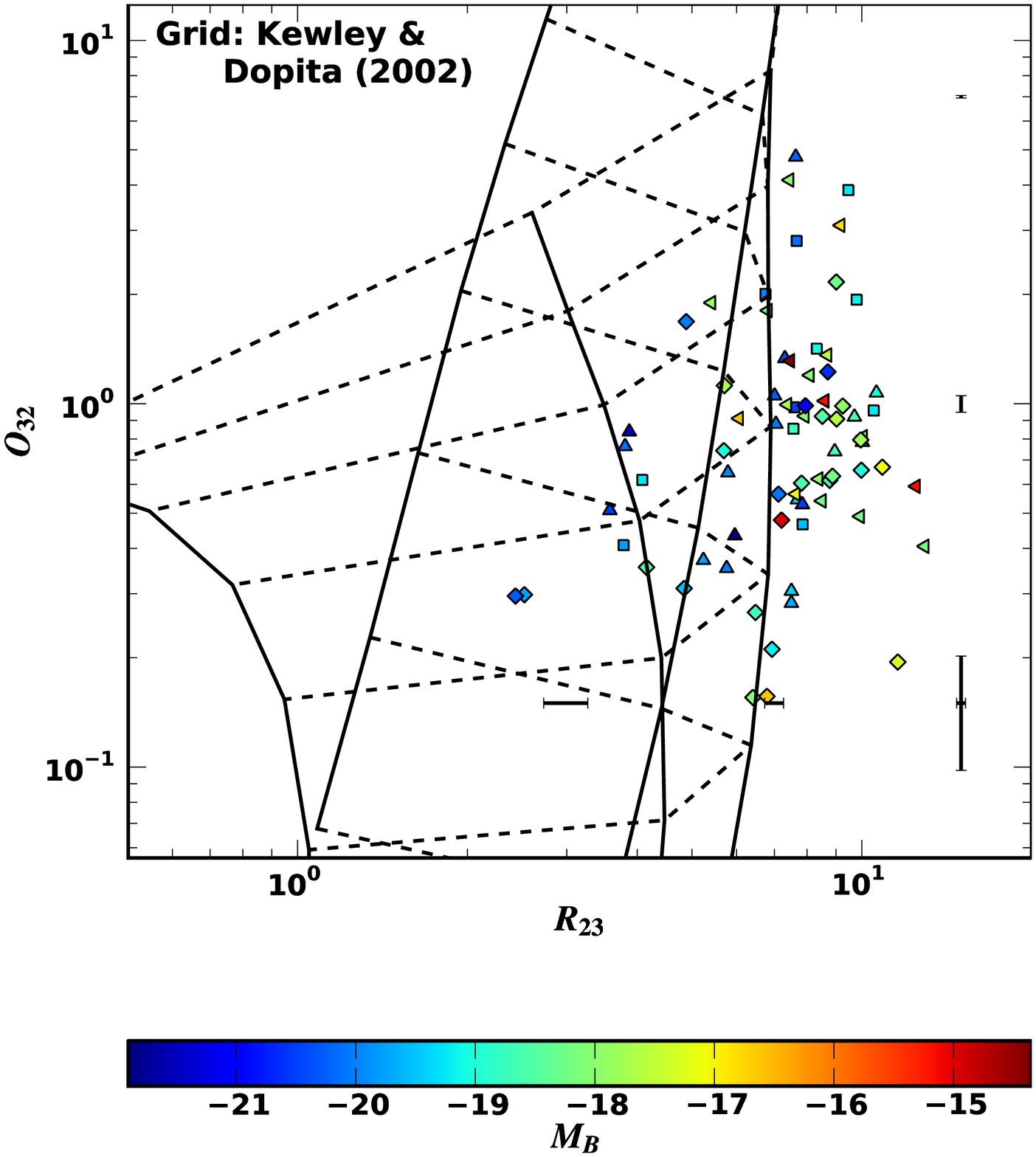}{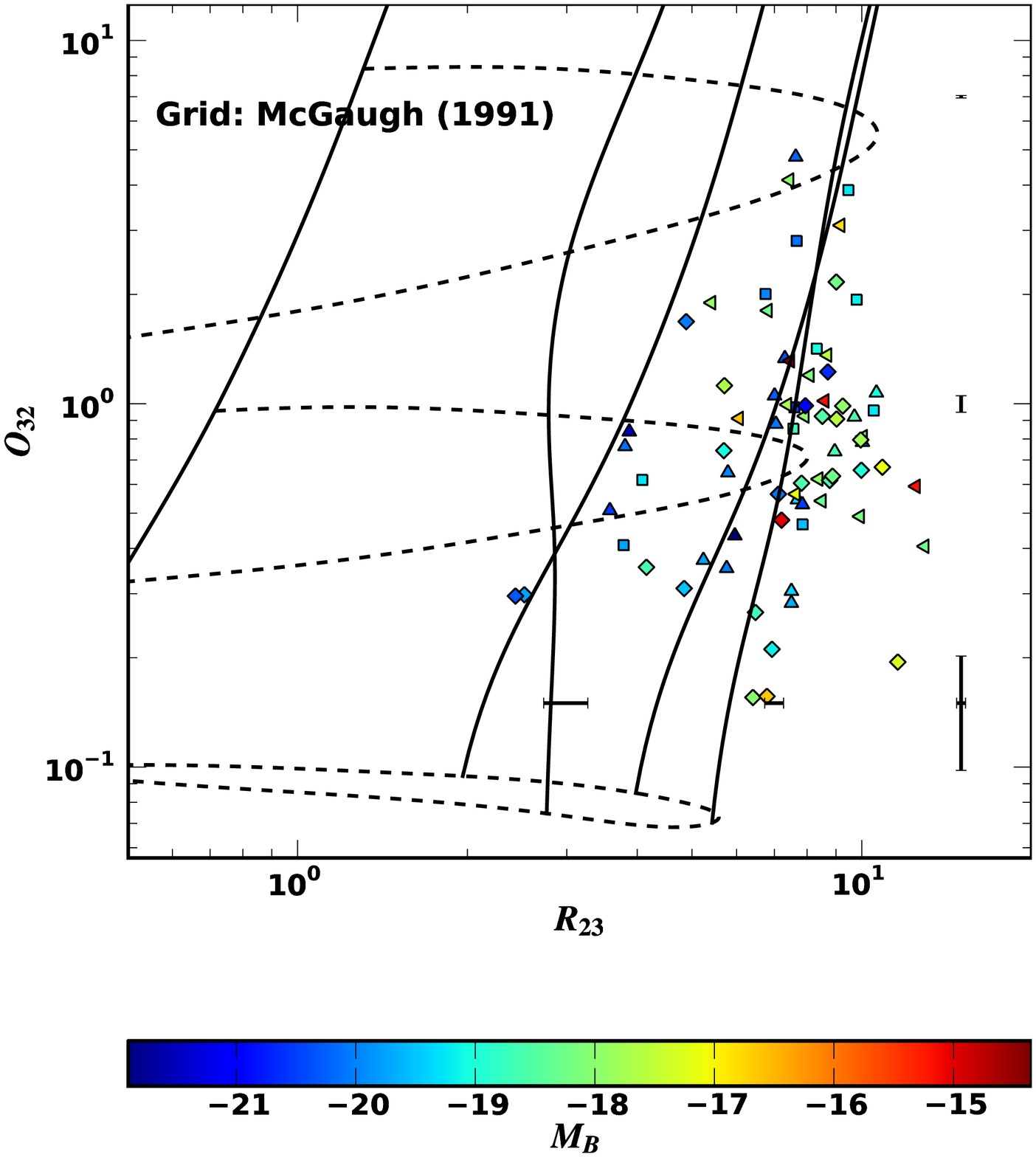}
\caption{$R_{23}$ vs. $O_{32}$ equivalent width ratios for  WHIQII and metallicity/ionization parameter grids.  Squares are LCBGs, up-pointing triangles are luminous and blue but not compact, left-pointing triangles are compact and blue but not luminous, right-pointing triangles are blue but neither compact nor luminous, and diamonds are the remainder. Color indicates absolute K-corrected B-band magnitude determined using spectroscopic redshifts. The left panel is the grid of \citet{KD02} with solid lines isometallicity contours ($\log(O/H)+12=9.1,8.6,8.1,7.9,7.6$ from lower-left around through upper-left), and dashed lines constant ionization parameter ($\log(U)= \log(q/c) = -2,-3,-4$ from top to bottom). The right panel shows the grid of \citet{MG91}, with solid lines isometallicity contours($\log(O/H)+12=9.0,8.7,8.3,8.0,7.4$ from lower-left around through upper-left), and dashed lines constant ionization parameter ($\log(q/c) = -2,-2.3,-2.6,-2.9,-3.2,-3.5,-3.8$ from top to bottom). Error bars along the lower-right edge indicate averaged errors. }
\label{fig:r23o32grids}
\end{figure*}

Two other issues arise in the use of $R_{23}$ beyond that of its uncertain calibration, both of which are apparent in the grid of Figure \ref{fig:r23o32grids} from \citet{KD02}.  First, the mapping between $R_{23}$ and oxygen abundance is complicated by the ionization parameter of the source HII regions.  The use of the $O_{32} = \oiii / \oii$ parameter circumvents this problem allowing for theoretically much more precise abundance measurements. While we always use both $R_{23}$ and $O_{32}$ in deriving oxygen abundances, we will continue to refer to the technique as simply $R_{23}$ for brevity. While it is not clear that the luminosity-weighted average of all HII regions in a galaxy should map to a ``global'' ionization parameter, the results of \citet{MK06abund} suggest that integrated line widths do  broadly follow the individual HII region metallicities for a variety of galaxy types and hence are viable for intermediate redshift galaxies where only integrated spectra are available.  Nevertheless, it is apparent that metallicity and ionization parameter are tied together in a non-trivial way that may vary in unusual galaxies in ways that photoionization models do not fully account for.

The second complication in the use of $R_{23}$ is the double-valued nature of the mapping from $R_{23}$ to oxygen abundance.  On the lower (metallicity) branch, the increasing oxygen abundances result in greater emission line strengths, causing  $R_{23}$ to increase with abundance.  At $\log(O/H)+12 \sim 8.3$, however, the increased cooling from emission lines begins to reduce the electron temperature enough to overcome this effect, causing   $R_{23}$ to now \emph{decrease} with abundance.  As a result, a second metallicity indicator is often necessary to break the degeneracy between these two branches.  Unfortunately, a second indicator is often not available at intermediate redshifts, forcing assumptions about the branch the galaxies lie upon.  

Because of the uncertainty associated with these model systematics and the nature of $R_{23}$, we again stress that the uncertainties in the $\log(O/H)+12$ values given in \ref{tab:specdata} are dominated by model systematics rather than the observational errors, but the relative values should be consistent\citep{KD08}.  Hence we only express the observational errors here, computed by directly calculating the metallicity for $1\sigma$ above and below the emission line equivalent width measurements.  As a further result of these uncertainties, we attempt to use the $R_{23}$ vs. $O_{32}$ plane directly as a guide to comparing samples, wherever possible.

The $R_{23}$ vs. $O_{32}$ relation for the WHIQII sample is shown in Figure \ref{fig:r23o32grids}.  Squares are LCBGs while other symbols are the remainder of the WHIQII spectroscopic sample.  Superimposed in the left panel is the grid \footnote{http://www.ifa.hawaii.edu/~kewley/Mappings/}  of \citet{KD02}, and in the right panel is the grid of \citet{MG91}\footnote{http://www.astro.umd.edu/~ssm/data/}.  Some of the WHIQII objects fall off the grids (by up to $\sim .2$ dex), an effect that has been noted before \citep[e.g.,][Fig. 2, Fig. 3]{MG94,KK04}.  It may  simply result from the problems with the photoionization codes as suggested in \citet{KD08}, or a violation of the assumption of an instantaneous burst \citep{KK04}.  Alternatively, because the grid shifts to the right for a top-heavy IMF with a large number of very high mass stars \citep{MG94}, this result may be an indicator of unusual stellar populations. 

\subsection{Mass-Metallicity Relation}
\label{sec:MZ}

\begin{figure*}[htbp!]
\plottwo{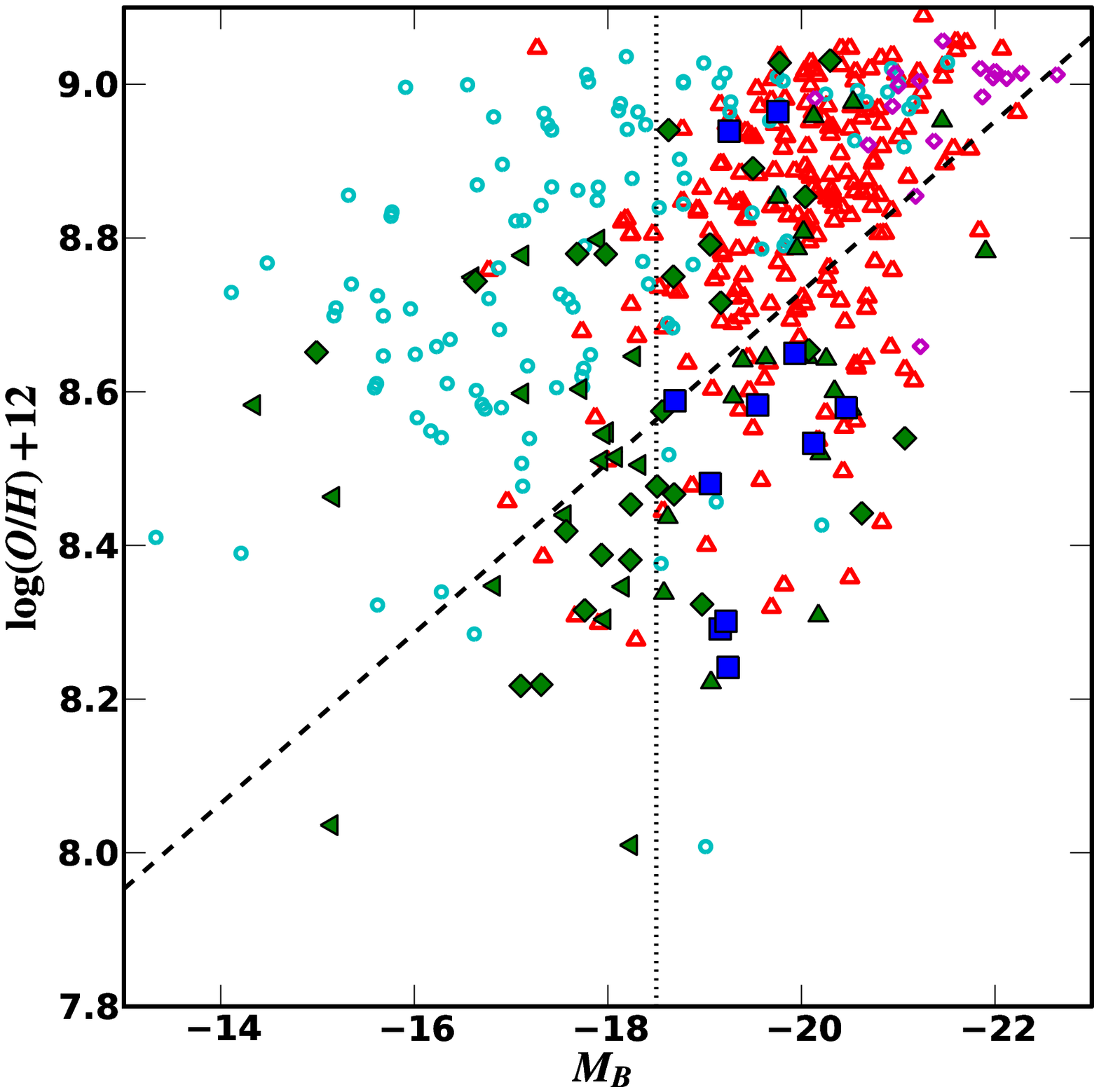}{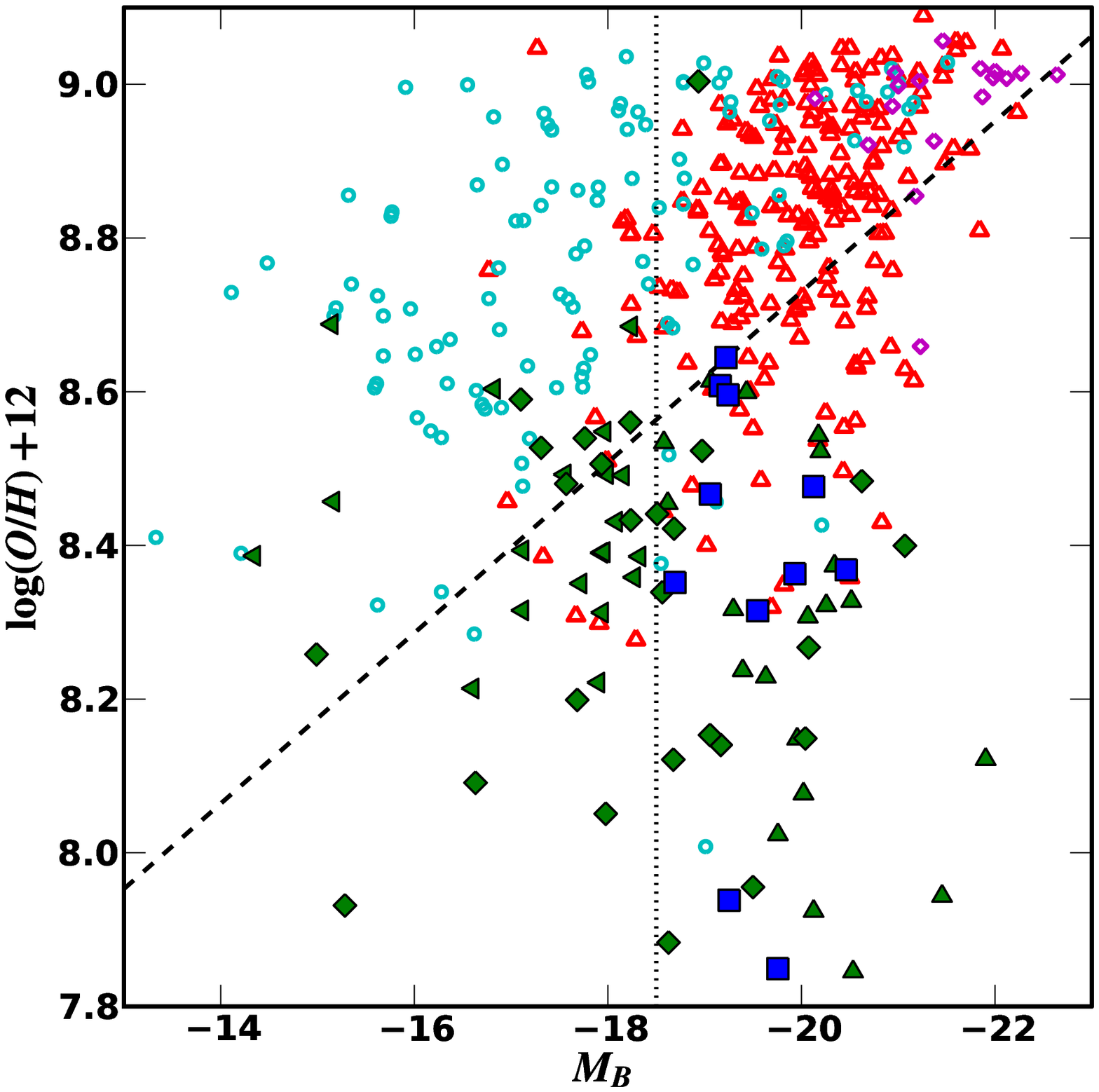}
\caption{LZ relation for WHIQII objects and comparison samples. WHIQII filled squares (blue) are LCBGs and other filled (green) symbols are non-LCBGs with the same symbol scheme as Figure \ref{fig:r23o32grids}.  The left panel assumes the galaxies lie on the upper branch of the $R_{23}$-$\log(O/H)$ relation, while the right panel is for the lower branch.   Open symbols are comparison samples described in detail in \S \ref{sec:comp}: open circles (cyan) are from \citet{nfgs00}, open diamonds (magenta) are \citet{CL01} and open triangles (red) are data points from \citet{KK04}.  The dashed line is the LZ relation for $0.4 < z < 0.6$ from \citet{KK04}. The vertical dotted line represents the lower limit for LCBG luminosities following our definition.}
\label{fig:lz}
\end{figure*}

Figure \ref{fig:lz} shows the luminosity-metallicity (LZ) relation for the WHIQII sample assuming the upper branch of the $R_{23}$-$\log(O/H+12)$, as well as comparison samples.   The comparison sample metallicities are derived by following the exact same analysis as that for the WHIQII objects using the equivalent width $R_{23}$ and $O_{32}$ values -- the points are recomputed from tabulated equivalent widths using the same calibration and methods.  The two panels illustrate the difference between use of the two $R_{23}$ branches for the WHIQII objects -- the left panel assumes the upper branch, and the right assumes the lower branch.  The \citet{KK04} line is derived from the mean of the $0.4 < z < 0.6$ and $0.6 < z < 0.8$ fits, which is approximately the range covered by the WHIQII galaxies. These figures validate the use of the upper branch  for most of the WHIQII sample; WHIQII objects on the lower branch fall $.5-1$ dex in metallicity below the LZ relation shown here, which uses the same calibration as our sample and is consistent with other LZ relations such as  \citealt{tremonti04} and \citealt{lamar04}.  This constrains all but a small fraction of WHIQII galaxies to lie on the upper branch.  For this reason and others discussed below, we assume the upper branch for all further analysis.

It is also clear in Figure \ref{fig:lz} that the locus of WHIQII points that are LCBGs or luminous and blue is shifted to systematically lower metallicity than the local samples, but is consistent with the $z \sim 0.5$ L-Z relation of \citet{KK04}.   Thus, \emph{metallicities of LCBGs are consistent with the typical intermediate redshift LZ relation} rather than that of some intrinsically unusual class.

\begin{figure}[htbp!]
\plotone{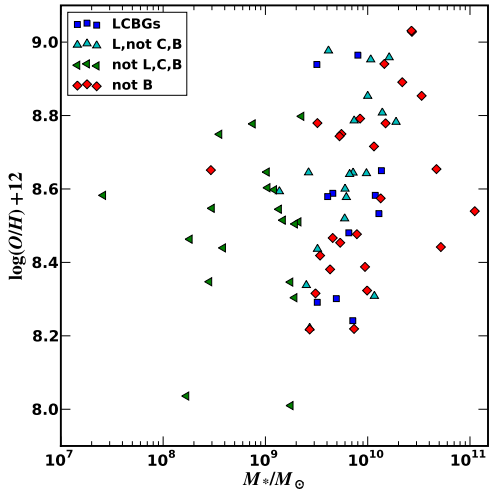}
\caption{$M_*-Z$ relation for WHIQII LCBGs (blue squares), luminous and blue but not compact objects (cyan up-pointing triangles), blue but not luminous objects (green left-pointing triangles), and objects that are not blue (red diamonds).  Stellar masses are derived from \citet{bdj01mstar}.}
\label{fig:mz}
\end{figure}

Finally, while there appears to be a noticeable offset between the luminous/blue  galaxies (inclusive of LCBGs) and the rest of the WHIQII spectroscopic sample, this is a selection effect due to the very blue colors and the use of $M_B$ instead of $M_*$. Figure \ref{fig:mz} shows the stellar mass-metallicity relation for the WHIQII objects -- the offset disappears and the populations are well mixed.  Furthermore, while the mass-metallicity relation also appears rather weak in Figure \ref{fig:mz}, it is apparent from Figure \ref{fig:lz} that this is simply a selection effect - the WHIQII galaxies are primarily selected from a relatively narrow luminosity range, and hence do not sample the galaxy population enough to show a strong LZ relation.

\begin{figure}[htbp!]
\plotone{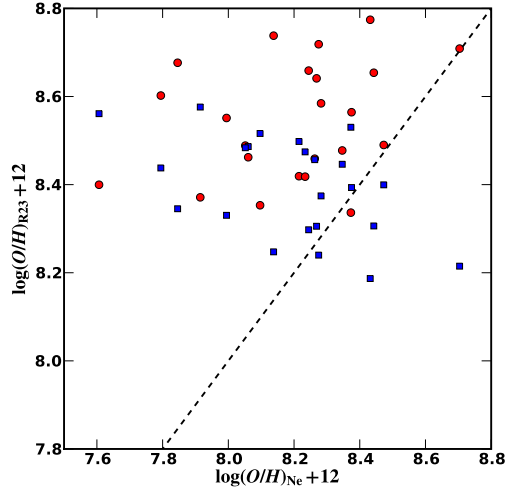}
\caption{Comparison of oxygen abundances derived from the Ne3O2 indicator (x-axis) vs. abundances from $R_{23}$ and $O_{32}$ with the \citet{KD02} calibration (y-axis) for the WHIQII sample with [NeIII] emission. Circles (red) indicate assuming the upper branch for $R_{23}$, and squares (blue) assume all objects are on the lower branch.}
\label{fig:ner23comp}
\end{figure}

Figure \ref{fig:ner23comp} compares the oxygen abundances from the Ne3O2 estimator to those from $R_{23}$ with either choice of branches for all WHIQII galaxies with [NeIII].  The disagreement in oxygen abundance between these techniques is large, preventing the use of Ne3O2 to select a branch.  The LZ relation apparent in Figure \ref{fig:lz} is not present if the Ne3O2 estimator is used, casting doubt on its utility for measuring metallicities  of intermediate redshift galaxies.

\section{Analysis}
\label{sec:disc}

\subsection{Compactness and Metallicity}

A central question in understanding the future of LCBGs is the nature of their compactness.  The scenarios outlined in \S \ref{sec:intro} make distinct predictions: if LCBGs are bursting dwarfs, they should have metallicities consistent with dwarf galaxies.  In fact, they might be systematically lower that local dwarfs, due to metallicity evolution. If they are bulges-in-formation, they should have metallicities more like those of normal spiral galaxy bulges.  In either case, the results of \citet{ell08mzsiz} show that, locally, compact galaxies tend to have higher metallicities than the mean LZ relation, so this might also be expected at higher redshift.  Hence, understanding how metallicity varies with compactness is important for unraveling the nature of LCBGs.

\begin{figure}[htbp!]
\plotone{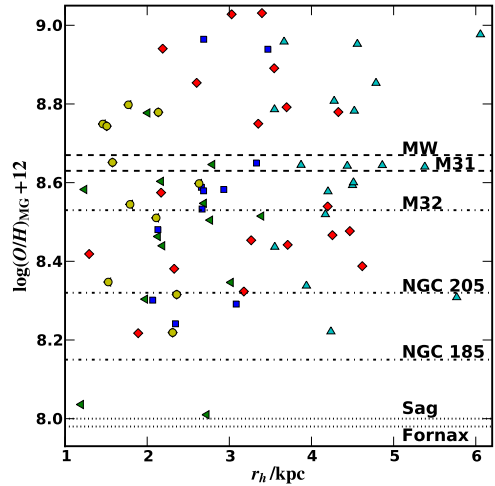}
\caption{Oxygen abundance measurements for WHIQII LCBGs  as a function of $r_h$. Squares (blue) are the WHIQII LCBGs (assumed to be on the upper branch for $R_{23}$), circles (yellow) correspond to objects at low enough redshift to measure $\niiha$ and hence break the $R_{23}$ degeneracy (note that none of these objects meet the LCBGs selection criteria).  Other symbols are non-LCBGs in WHIQII following the same symbols as in Figure \ref{fig:mz}  Horizontal lines are the \citet{richer98} metallicities from PNe of Local Group hot stellar systems. Dashed lines are the M31 and Milky Way bulges (from top to bottom) dashed-dotted lines are bright M31 satellites (NGC 205, M32, NGC 185) and dotted lines are MW satellites (Sagittarius and Fornax). }
\label{fig:rhoh}
\end{figure}

Figure \ref{fig:rhoh} shows the derived oxygen abundance as a function of $r_h$ for the spectroscopic sample of WHIQII.  None of the intermediate redshift objects have detectable emission suitable for breaking the $R_{23}$ degeneracy, but we also show local targets with detectable $[NII]$ and $H\alpha$.  Most of these galaxies lie on the upper branch.  They share at least some of the selection criteria of the LCBGs, offering more evidence that the LCBGs primarily lie on the upper branch, as expected from the luminosity-metallicity relation (\S \ref{sec:samp} and \S \ref{sec:MZ}).  Figure \ref{fig:rhoh} also shows a result that is quite surprising: there is at best a weak correlation between compactness and oxygen abundance.  
To better quantify the amount of correlation, we perform Monte Carlo simulations where we offset the data points based on the observational error bars in both $r_h$ (from the model as described in \S \ref{sec:dat}) and $\log(O/H)+12$, and we bootstrap resample the data set.  We then calculate the standard Pearson product-moment correlation coefficient.  Fitting a Gaussian to this distribution shows that the correlation coefficient for this relation is within $0.9 \sigma$ of 0 for the sample of luminous and blue objects, conservatively assuming no intrinsic scatter to the relation.  We also consider how much of a correlation could be hidden by these errors.  We find that any linear relation passing through our points of the form $r_h = m (\log(O/H))+b$ must have a slope of at least $m=.11$ to be detected at the $3 \sigma$ level using the significance test described above, and the best-fit slope for our data set is $m=0.6$.  Hence this relation is at best consistent with a weak tendency for bigger galaxies to be more metal rich, and reasonably consistent with no correlation at all.  

This result is apparently in contrast to the local result of \citet{ell08mzsiz} that galaxies in the SDSS with smaller $r_h$ have high metallicities.  If an $r_h$ vs. $\log(O/H)+12$ trend comparable to that seen for the  range $9 < M_* < 10$ in \citet{ell08mzsiz} Figure 2 were present with minimal intrinsic scatter we would detect it at roughly the $3\sigma$ level based on the discussion above.  However, the high star formation rates of WHIQII galaxies as well the higher redshift of this sample as compared to SDSS imply that the specific star formation rates (SSFR) should be higher \citep[e.g.][]{lehn09}.  This would offset the $r_h$ trend, as \citet{ell08mzsiz} also detected a bias of lower metallicity for higher SSFR.  While the noted trend for this $M_*$ range would not necessarily completely cancel out the $r_h$ trend, it would likely reduce it below the level of detectability and/or add more scatter.  Hence, the WHIQII result could well be consistent with \citet{ell08mzsiz} within the errors.  Furthermore, as discussed below in \S \ref{sec:comp}, there are signs of redshift evolution in the WHIQII sample that explain the generally lower metallicities than the SDSS \citep{ell08mzsiz}.  Most of the potential mechanisms behind the correlations are likely to evolve in that time, so it is not at all clear that the \citet{ell08mzsiz} result applies to intermediate redshift.

\begin{figure}[htbp!]
\plotone{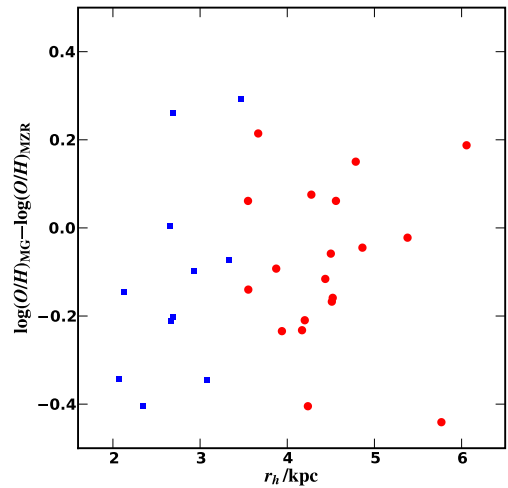}
\caption{Oxygen abundance measurement residuals from LZ relations for WHIQII LCBGs as a function of $r_h$. Residuals shown are against the LZ of \citet{KK04}.  We plot luminous and blue galaxies that satisfy the luminosity cut as blue squares, and those that do not as red circles}
\label{fig:rhohres}
\end{figure}

\begin{figure}[htbp!]
\plotone{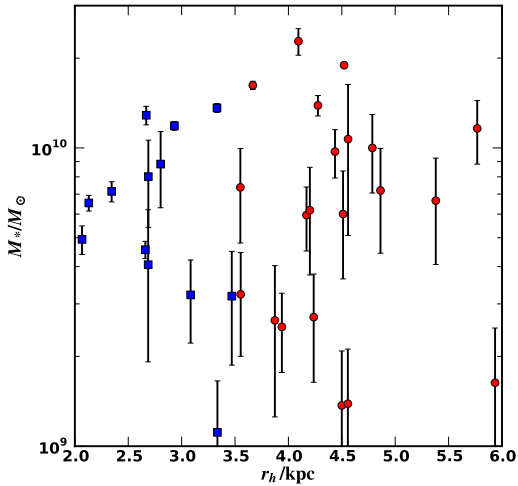}
\caption{\citet{bdj01mstar} stellar mass for WHIQII objects that meet the luminosity and color criterion of LCBGs as a function of $r_h$. Blue squares have $r_h<3.5$, red circles do not.}
\label{fig:rhvsmstar}
\end{figure}

Naively, either of the two evolutionary scenarios discussed above would be expected to show a relationship between $r_h$ and oxygen abundance -- dwarf galaxies and bulges-in-formation should obey distinct relations given their systematically different locations along the LZ relation.  Nevertheless, from Figure \ref{fig:rhoh} it is clear that this trend is weak or not-existent -- the apparent size of starbursts cannot be used to predict their metallicity.   The weak correlation in Figure \ref{fig:rhoh} may be a result of a combination of the L-Z relation and an L-$r_h$ relation, rather than the direct correlation between $r_h$ and Z that we are seeking here.  This effect is mitigated by the fact that we are sampling a relatively small range in the LZ relation (see \S \ref{sec:MZ}).  To search for it, in Figure \ref{fig:rhohres} we show the \emph{residuals} against the LZ relation of \citet{KK04}.   Again, there is no apparent correlation for the residuals.  Repeating the correlation coefficient analysis discussed above for the residuals against the LZ relation of \citet{KK04} discussed in \S \ref{sec:met} (it uses the same calibration we use here)  shows that the relation is within $0.8\sigma$ of a coefficient of 0. This result shows that there is even less detectable correlation than in the direct $r_h$-Z relation. In Figure \ref{fig:rhvsmstar} we show stellar mass plotted against $r_h$ for the WHIQII LCBGs.  The same lack of correlation appears here -- more compact objects do not necessarily have lower stellar mass, and the correlation is within $0.1\sigma$ of 0.  These results make clear that \emph{compactness of a starburst at intermediate redshift does not strongly correlate with present epoch properties} such as stellar mass or metallicity.

\subsection{Evolution}
\label{sec:comp}

\begin{figure*}[htbp!]
\plottwo{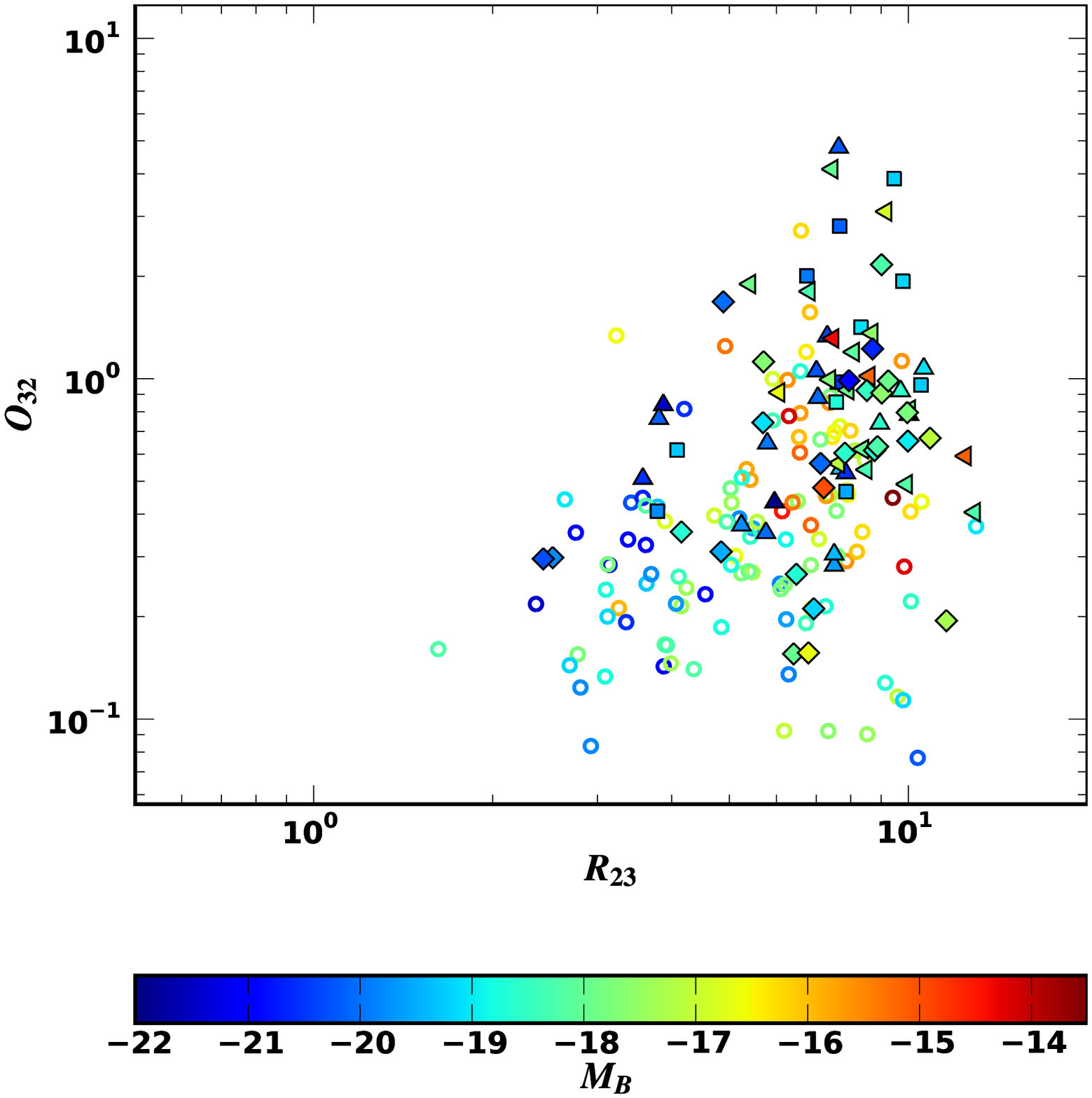}{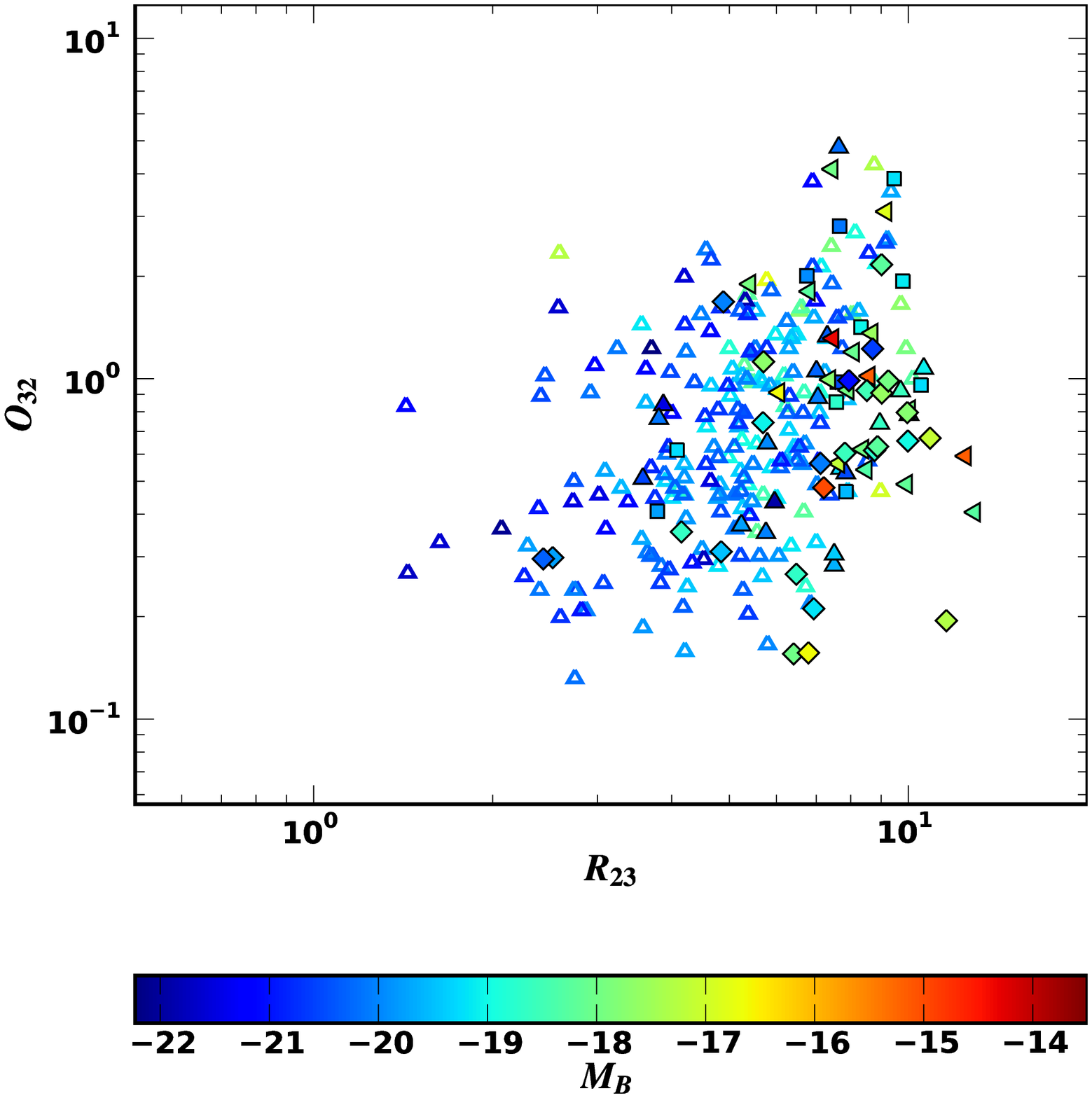}
\caption{$R_{23}$ vs. $O_{32}$ equivalent width ratios for  WHIQII (filled symbols) and selected comparison samples (open symbols).  The filled symbols follow the pattern of Figure \ref{fig:r23o32grids}, while the left panel has  the local survey of \citet{nfgs00} as open circles and the right panel shows the intermediate redshift sample of \citet{KK04} as open triangles.  Color indicates absolute B-band magnitude, although the scale is not the same as Figure \ref{fig:r23o32grids} due to a larger dynamic range in the comparison samples. }
\label{fig:r23o32multi}
\end{figure*}

In the left panel of Figure \ref{fig:lz} we compare the WHIQII survey LZ relation to the Nearby Field Galaxy Survey of \citet[][NFGS]{nfgs00} and the intermediate redshift sample of \citet{KK04}.  There is a distinct offset between the locus of points occupied by WHIQII galaxies and that of \citet{nfgs00} -- the WHIQII samples is systematically lower in metallicity.  This trend also appears when comparing Figure \ref{fig:mz} to the corresponding figures of \citet[][Figures 1 and 2]{ell08mzsiz} -- the WHIQII sample is distinctly lower in metallicity even compared to the high $r_h$ or low SSFR trends.  In Figure \ref{fig:r23o32multi} we compare  WHIQII to the same two comparison samples in the $R_{23}$/$O_{32}$ plane.  Our calculations use equivalent width ratios to ensure a fair comparison and to match the grids of Figure \ref{fig:r23o32grids}.    There is a distinct offset of $\sim .5 \pm .3$ dex between the locus of the WHIQII luminous/blue objects and this local sample objects of comparable magnitude.  These differences may be fundamental to the differences between LCBGs and the local galaxy population.  It is apparent in Figure \ref{fig:r23o32multi} that the NFGS galaxies lie lower and to the left than the luminous/blue WHIQII objects and hence are higher metallicity and likely also lower ionization parameter.  Qualitatively similar results appear for the local sample of \citet[][not shown]{MK06cat}.  In contrast, the intermediate redshift sample of \citet{KK04} is more consistent with the WHIQII galaxies (mean $\sim .2 \pm .3$ dex), although there are significantly more very high metallicity objects, likely due to the broader redshift range and the lower fraction of compact objects in \citet{KK04} relative to WHIQII.  Overall, for a given luminosity, the intermediate redshift sample is better matched to the WHIQII sample than the locus of the same luminosity in the local sample.

To more directly address the question of what broad classes of objects are the present epoch end states of LCBGs, we consider some of the candidate local objects.  To this end, we compare the metallicities of the WHIQII LCBGs to a variety of different comparison samples:  Local Group galaxies (especially dwarfs) as determined by planetary nebulae, HII regions in dwarf irregular galaxies, and nuclear starbursts.  Due to the uncertainties surrounding the calibration of the $R_{23}$ estimator of oxygen abundance, even in cases where the gas phase nebular lines \emph{are} available, we rely on empirical comparison in the $R_{23}$/$O_{32}$ plane as a way to compare between samples.  Future comparisons may benefit from  direct use of $R_{23}$ and $O_{32}$ (or the emission line fluxes).

Quiescent dwarf galaxies in the Local Group like the dwarf elliptical brighter companions of M31 (NGC 205, NGC 185, M32), or the brightest Milky Way dwarf spheroidals, have been suggested as possible end points of LCBGs \citep[e.g.,][]{koo95,guz98,crawford06,hoyos07}.  Directly comparing $R_{23}$ and $O_{32}$ is impossible for these objects, as they lack nebular emission lines.  However, planetary nebulae (PNe) can provide an estimate of the (stellar) oxygen abundances using similar techniques to those used for HII region emission, although they exhibit the higher ionization and densities found in PN.  While comparing PN (evolved stellar) oxygen abundances to the gas phase is not ideal, it is the only way available to determine oxygen abundances in these passive stellar systems.  Given the fact that the scatter of a galaxy's PN-derived oxygen abundance is roughly the same as the uncertainty in the $R_{23}$ calibrations, it may represent a reasonable first-order comparison.  \citet{richer98} summarize results from such observations, as Figure \ref{fig:rhoh} shows (horizontal lines).   Unless a sizable fraction of the galaxies lie on the lower branch, which is unlikely by arguments presented above, most are already too high metallicity to fade into even the brightest Milky Way dSphs. If, instead, most lie on the upper branch, half are still too metal rich to fade into even NGC 205, which has the highest oxygen abundance of the dwarfs in the sample.  A few are even high enough to be consistent with $L_*$ spiral bulges, as suggested by the abundances of the Milky Way and M31 bulges.  This argument supports the notion that LCBGs are a heterogeneous population, where some will evolve into bulges, and others into quiescent sub-$L_*$ galaxies.  Our most surprising result is that the physical size of the starburst does not predict which of these scenarios will occur.

\begin{figure}[htbp!]
\plotone{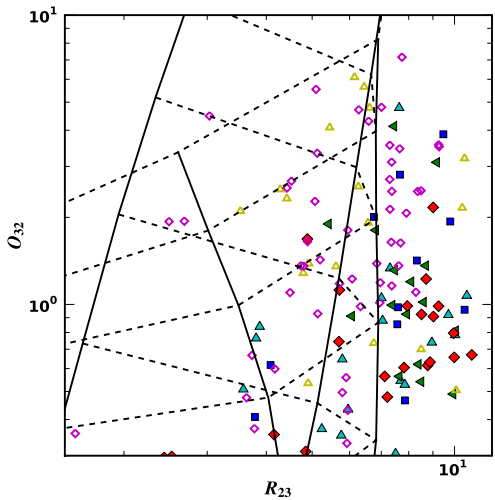}
\caption{$R_{23}$ vs. $O_{32}$ for WHIQII LCBGs (filled blue squares), non-compact LCBGs (other filled symbols following Figure \ref{fig:mz}), as well as dIrrs from \citet[][yellow open triangles]{kuz04} and \citet[][magenta open diamonds]{vzh06}.  The grid is that of \citet{KD02}, also shown in Figure \ref{fig:r23o32grids} (upper panel) for comparison. }
\label{fig:roirr}
\end{figure}

\begin{figure}[htbp!]
\plotone{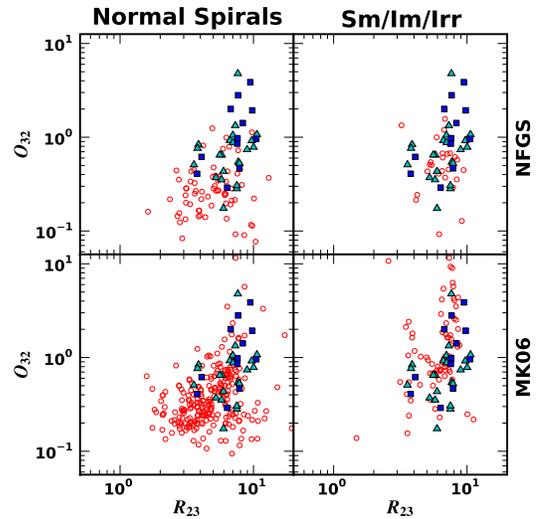}
\caption{$R_{23}$ vs. $O_{32}$ comparison between WHIQII and selected local surveys.  In each panel, squares (blue) are WHIQII LCBGs, triangles (cyan) are WHIQII objects that are luminous and blue but not compact, while open circles (red) are the comparison sample.  The upper-left panel comparison sample is derived from NFGS \citep{nfgs00} galaxies that are categorized as normal spirals (i.e. not E,S0,Sm,or I), while upper-right are irregulars (Sm and Im) from the same survey.  Lower-left are spirals from \citet{MK06cat}, and lower-right are those categorized as irregulars ( Im or irr).}
\label{fig:rocompsirr}
\end{figure}

Another potential comparison sample is that of dwarf irregular (dIrr) galaxies or other star-forming blue galaxies \citep[e.g.,][]{guz97,Garland07,hoyos07}.  In the local universe, these galaxies are fairly common, and are resolved well enough to obtain spectra of the individual HII regions that combine in a luminosity-weighted fashion to give the integrated emission lines measurable at intermediate redshift.  Figure \ref{fig:roirr} shows the WHIQII LCBGs, as well as objects that are luminous and blue but not compact enough, although as shown in Section \ref{sec:samp}, this sample is still much more compact than the general field population.  We compare to two such HII region samples \citep{kuz04,vzh06} on the $R_{23}$,$O_{32}$ diagram.  They show offsets in the mean of $.2 \pm .3$ and $.1 \pm .3$, for \citet{kuz04} and \citet{vzh06}, respectively.  From this analysis, it is apparent that the WHIQII LCBGs occupy a similar region in this diagram,  at least in contrast to the general sample of local galaxies shown in Figures \ref{fig:r23o32multi}.   Furthermore, Figure \ref{fig:rocompsirr} plots the WHIQII objects in the same plane comparing normal spirals in the integrated spectra of \citet{nfgs00} and \citet{MK06cat} to irregulars.  The WHIQII objects are preferentially in the same region, but there is significant overlap between parts of the spiral population (generally lower-luminosity) and WHIQII.  This suggests that there is some connection between the star formation modes present in LCBGs and those of irregulars, perhaps a higher density of lower-metallicity star formation, or localized instead of global modes. 

\begin{figure}[htbp!]
\plotone{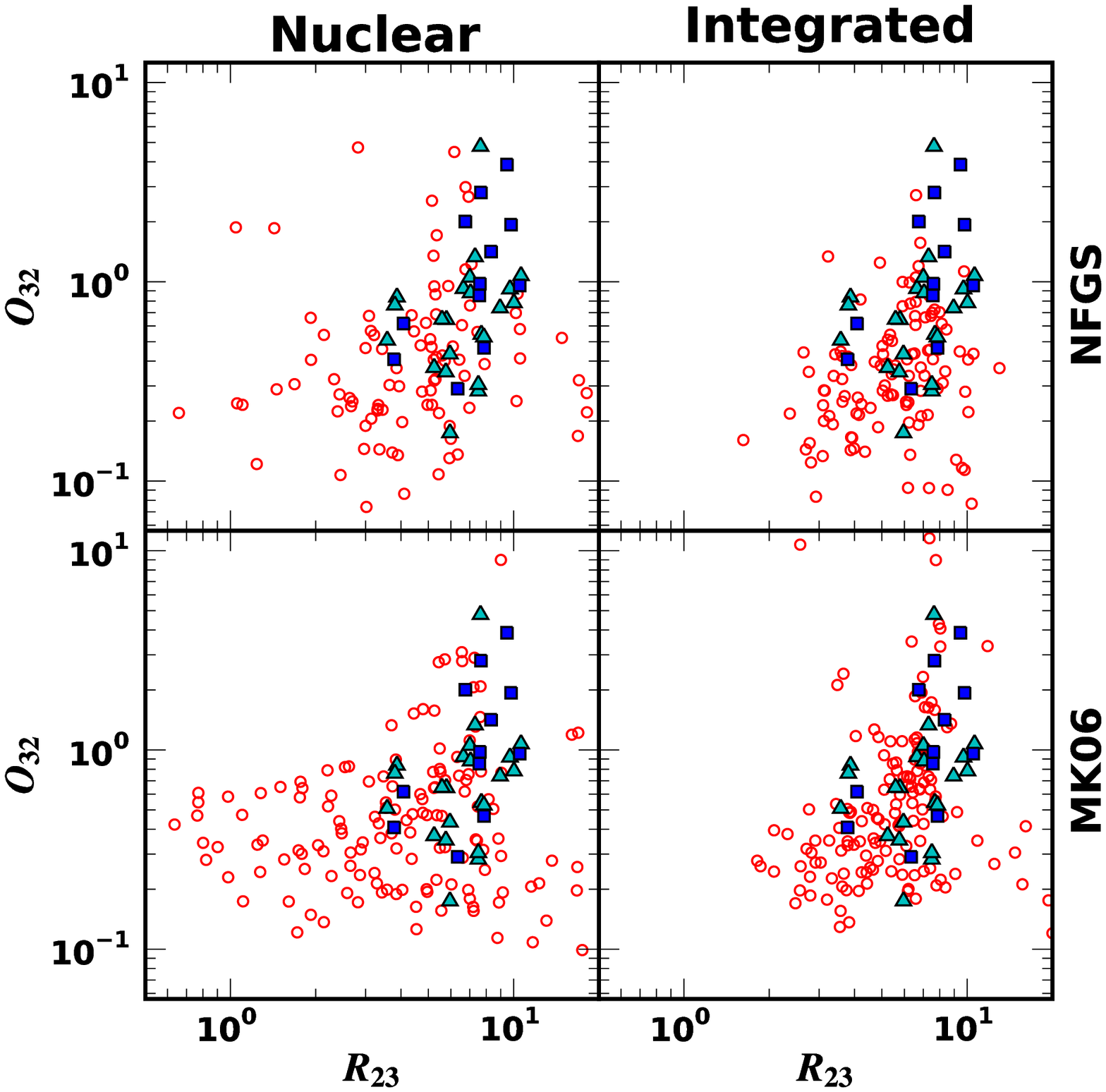}
\caption{$R_{23}$ vs. $O_{32}$ comparison between WHIQII and selected local surveys.  In each panel, squares (blue) are WHIQII LCBGs, triangles (cyan) are WHIQII objects that are luminous and blue but not compact, while circles (red) are the comparison sample.  The upper-left panel comparison sample is derived from the nuclear spectra of the NFGS \citep{nfgs00}, while upper-right is from the integrated spectra of the same survey.  Lower-left is nuclear emission from \citet{MK06cat}, and lower-right is the corresponding integrated ratios.}
\label{fig:rocompni}
\end{figure}

Nuclear star-forming regions may also provide a connection to LCBGs, as they are highly compact areas of intense star formation that can be studied relatively easily in the local universe by comparing spectra taken of nuclear regions of a well-resolved galaxy and comparing to a larger integrated-light spectrum of the whole galaxy.  Hence, in Figure \ref{fig:rocompni} we compare WHIQII objects to local samples that include both nuclear and integrated spectra for a large galaxy sample on the $R_{23}$ vs. $O_{32}$ plane.  While there is substantial overlap between the WHIQII points and both the nuclear and integrated spectra  of NFGS and \citet{MK06cat}, there seem to be a number of WHIQII objects (particularly true LCBGs) in the high $O_{32}$ region of the plane where the nuclear spectra are more coincident than the integrated spectra.   While this trend is somewhat weak, it does suggest a possible connection between WHIQII LCBGs and nuclear starbursts, or at least similar ionization parameter and metallicity properties.

The connection between LCBGs and irregular galaxies or local nuclear starbursts yields insights into the nature of the star formation and explains why the LCBGs are in an unusual locus compared to local counterparts in Figure \ref{fig:r23o32multi}.  Irregular and LCBGs both show higher ionization parameters than typical star forming galaxies.  Furthermore, some Irregulars lie off the grids of Figure \ref{fig:r23o32grids}, just as WHIQII LCBGs do.  This suggests that the star formation in LCBGs is analogous to irregular galaxies, with unusually intense star formation in a few very bright HII regions.  In the case of those LCBGs that are more consistent with low-luminosity spirals, a high ionization due to a high star formation rate in a relatively compact bulge also explains these same observations, particularly given the lower metallicities characteristic of higher redshifts.  Hence, the LCBG locus on the $R_{23}/O_{32}$ plane could be due to high ionization parameters from the intense star formation in a small volume that is essentially the very definition of the class.  This conclusion is further borne out in the fact that the entire WHIQII spectroscopic  sample in Figure \ref{fig:r23o32grids} -- not just the LCBGs -- also lies in the region of high ionization parameter.  The choice of sample inherently biases the population towards high-ionization objects by selecting objects that are luminous, compact, or blue relative to other objects, even if they are not necessarily bona fide LCBGs.   This is further enhanced by the higher star formation rate densities and pressures encountered at higher redshift \citep{liushapley08,hain09highzsf} which likely lead to higher ionization parameters\citep{brin08,lehn09}, and may render typical assumptions of locally-calibrated photoionization models inaccurate. While there is no strong trend in the $r_h$ vs. ionization parameter plane, it is not clear that this trend should exist --  WHIQII objects are mostly compact enough that they may all have relatively high ionization parameter. \citet{Z94} notes a similar lack of correlation, although their result is for individual HII regions vs. galactocentric distance rather than size of the overall galaxy.     

Most of the arguments above apply, to a somewhat varying extent, to \emph{all} intermediate redshift star-forming galaxies.  The intermediate redshift ``field'' sample of \citet{KK04} does sample a similar part of parameter space as WHIQII, although WHIQII is biased by the selection criteria (\S \ref{sec:isamp}, Figure \ref{fig:rhdist}).   In fact, LCBGs are objects that are only slightly more luminous and compact than average at intermediate redshift, as shown by our photometric sample (\S \ref{sec:numden}).  If so, their apparently strong evolution with redshift is just another example of ``downsizing,'' the shift in the characteristic scale of star formation as a function of redshift.  In fact, the location of such intermediate redshift objects in metallicity diagnostic diagrams tend to be intermediate between the local samples discussed here and high-redshift samples \citep[e.g.][]{erb06,liushapley08,lehn09,hain09highzsf},  suggesting just such a continuous process.

More exotic possibilities such as a top-heavy IMF \citep{MG94} in all of these objects cannot be excluded, but given the significant fraction of the total galaxy population that these objects represent, it is difficult to come up with a scenario where they are all extreme outliers. 

\subsection{Discussion}
We now consider some  important caveats to these results. One concern is the potential effect of low-luminosity AGN \citep[e.g.,][]{ho97,kewley01,kauf03}.  Because LCBGs are selected to be compact, selection effects would tend to increase the AGN fraction.  While we have eliminated the obvious cases (i.e. Figure \ref{fig:lamar}), there is the possibility that some of our objects are partially contaminated by AGN that are not bright enough to overpower the starburst, but nevertheless contribute somewhat to the $R_{23}$ and $O_{32}$ ratios.  This may be especially so at higher redshift \citep[e.g.,][for $z \sim 2$]{brin08,lehn09}

Despite our arguments to the contrary, the possibility exists that some substantial fraction of LCBGs may lie on the lower branch of $R_{23}$ and hence we may be assigning incorrect metallicities by choosing the upper branch. We note that Figure \ref{fig:r23o32grids} indicates that many of the WHIQII objects are near the turnover from one branch to the other. Those for which this is the case will not fall extraordinarily far from the upper branch relation, however -- near the turnover, the relation is close to symmetric.  Hence, this concern may be represented as an increase in the metallicity error associated with points near the turnover that are not included in the random errors of Figure \ref{fig:rhoh}.  Yet this ambiguity is small enough that all the results of the previous section still hold, as they apply to the whole set of objects rather than being sensitive to individual objects near the turnover.

\citet{hoyos05} notes the existence of a population of objects with very low metallicity that are still consistent with LCBG criteria in the DEEP2 \citep{davis03deep2} survey and the Team Keck Redshift Survey \citep[][ TKRS]{wirth04tkrs}.  While there may be objects of this type in the WHIQII LCBG sample, the objects with confirmed low abundances in the \citet{hoyos05} sample represent only $0.3 \%$ of the total TKRS+DEEP2 sample. Hence, it is very unlikely that more than $\sim 1$ of these objects has found its way into the WHIQII LCBG sample, which represents 10\%-20\% of galaxies at intermediate redshift.  Furthermore, the \emph{non}-detection of the $[OIII]\lambda 4363$ auroral line in any of the WHIQII LCBGs further strengthens the case that most of these objects lie on the upper branch -- the $R_{23}/O_{32}$-implied metallicities on the lower branch are easily low enough for some of the WHIQII LCBGs that at least a weak detection of the $[OIII]\lambda 4363$ line should have been possible for the highest SNR spectra.  

In the opposite sense, there are also low mass, high metallicity objects discussed in the literature. \citet{peep08} investigates these in some detail, noting that while these are extreme outliers to the $M_*$-metallicity relation, they nevertheless look like ordinary dwarf elliptical galaxies.  A few of these objects even show blue colors in their centers but are redder and fainter further out.  It is possible that objects like these are analogous to LCBGs, and while they are outliers in the present epoch, it is possible that they were much more common at intermediate redshift. However, given that the examples in \citet{peep08} are $0.04 \%$ of the population from \citet{tremonti04}, this population is unlikely to evolve quickly enough to represent a significant fraction of the $\sim 20 \%$ of galaxies that LCBGs represent at intermediate redshift. 

A final issue is that of the influence of gas flows.  \citet{erb06} report a high amount of gas outflow at high redshift at all galaxy masses (including the stellar masses included in the WHIQII LCBGs), while locally outflows are prevalent only for lower mass galaxies.  Where LCBGs fit into this picture is unclear - they may in fact have a higher fraction of bulges-in-formation than predicted from the metallicity results described above if they are characterized by anomalously strong outflows.  Furthermore, as shown in \citet{kew06}, galaxy interactions that trigger star formation can funnel gas to the center and produce an apparently lower metallicity due to infall of pristine gas.  The wide scatter of metallicities apparent in the WHIQII sample can neither prove nor disprove these scenarios as predominant, although it may suggest that gas flows can be important in intermediate redshift starbursts.

\section{Conclusions}
\label{sec:conc}
In this paper, we have analyzed the LCBGs found in the WHIQII survey to constrain their evolutionary path to the current epoch.     

\begin{enumerate}
\item LCBGs are likely a heterogeneous population.  The range of metallicities is too large to be consistent with an all-dwarf or all-spiral scenario, and some of the galaxies show metallicities simply too high to evolve into a typical present day dwarf.  Combined with other results like those of \citet{ham01, Garland04, bvz06cnelg, noeske06}, this study suggests that they are a mix of true dwarfs and in situ bulges forming in lower-luminosity spirals.  The WHIQII data effectively rule out either an all-dwarf or all-bulge descendant population.
\item WHIQII objects, including LCBGs, show signs of redshift evolution in the luminosity-metallicity relation.  They have lower metallicities at fixed stellar mass or luminosity, consistent with the results of \citet{KK04}.
\item Compact star forming galaxies have no strong correlation between size, $r_h$, and oxygen abundance or $M_*$.  That is, apparent size does not appear to correlate with the likely end product. Thus, apparently compact objects at intermediate redshift are not necessarily intrinsically compact or low mass galaxies.
\item LCBGs appear to have high ionization parameters relative to local galaxies, consistent with the intermediate redshift field \citep{KK04}.  While this result may be physically related to their compactness and intense star formation, it is also true of other field galaxies at intermediate redshift.  Hence, the observed evolution of LCBGs may be generic ``downsizing'' enhanced by selection effects.  
\end{enumerate}

If LCBGs are indeed a heterogeneous population, it may be that unusually blue galaxies seem uncommon simply because the \emph{global} star formation rate declines, more or less irrespective of star-forming galaxy type \citep{noe07,cw08}.  Hence, the conditions in their star forming regions would be intermediate between local galaxies and high-redshift objects such as Lyman Break Galaxies \citep{steidel98lbgs,shap01lbgs,shap03lbgs}.  This is one of the many faces of galaxy ``downsizing,'' and the WHIQII data appear to be consistent with this scenario.  Hence, while LCBGs are a straightforward category to observationally define, the scatter in their properties means that they must be individually examined to understand their place in the wider context of galaxy formation.

\vspace{10mm}
We wish to thank Lisa Kewley for extensive and insightful comments; we also thank David Koo, Janice Lee, Alice Shapley, Marla Geha, and the anonymous referee for conversation and helpful suggestions, as well as J.D. Smith for assistance with the imaging observations. This research was supported by NSF grant AST05-06054 and the Center for Cosmology at UC Irvine.  The authors wish to recognize and acknowledge the very significant cultural role and reverence that the summit of Mauna Kea has always had within the indigenous Hawaiian community.  We are most fortunate to have the opportunity to conduct observations from this mountain. 

{\it Facilities:} \facility{WIYN (MiniMo)}, \facility{Keck:I (LRIS)}

\bibliography{ms}{}
\bibliographystyle{hapj} 

\clearpage
\begin{deluxetable}{cccccccc}
\tabletypesize{\scriptsize}

\tablecolumns{8}
\tablecaption{Photometric properties of WHIQII objects with redshifts.}
\tablehead{
\colhead{WHIQII \#\tablenotemark{a}} &
\colhead{RA\tablenotemark{b}} &
\colhead{Dec\tablenotemark{c}} &
\colhead{$B$\tablenotemark{d}} &
\colhead{$B-V$\tablenotemark{e}} &
\colhead{$M_{star}/M_{\odot}$\tablenotemark{f}}  &
\colhead{$r_h/{\rm kpc}$\tablenotemark{g}} &
\colhead{$z$\tablenotemark{h}} 
}

\startdata
26114 & 177.67011 & 28.754773 & $-20.2 \pm 0.2$ & $0.5 \pm 0.3$ & $1.42 \times 10^{10}$ & 3.5 & $0.6387 \pm 0.0005$ \\
81835 & 327.64159 & 28.851717 & $-16.8 \pm 0.2$ & $1.0 \pm 0.3$ & $5.75 \times 10^{9}$ & 2.1 & $0.2118 \pm 0.0001$ \\
24584 & 177.63101 & 28.787589 & $-21.9 \pm 0.1$ & $0.2 \pm 0.3$ & $4.82 \times 10^{9}$ & 8.4 & $0.7795 \pm 0.0004$ \\
34826 & 218.84771 & 24.887107 & $-16.7 \pm 0.1$ & $0.5 \pm 0.2$ & $5.67 \times 10^{8}$ & 3.0 & $0.1703 \pm 0.0000$ \\
82434 & 327.66416 & 28.860294 & $-18.5 \pm 0.2$ & $0.7 \pm 0.3$ & $7.84 \times 10^{9}$ & 4.5 & $0.3910 \pm 0.0004$ \\
50703 & 258.04971 & 33.554882 & $-20.7 \pm 0.1$ & $0.5 \pm 0.2$ & $2.28 \times 10^{10}$ & 4.1 & $0.4875 \pm 0.0002$ \\
42512 & 58.880567 & 9.7606045 & $-21.1 \pm 0.2$ & $0.6 \pm 0.3$ & $4.65 \times 10^{10}$ & 6.2 & $0.8015 \pm 0.0000$ \\
41816 & 58.862955 & 9.7311169 & $-21.2 \pm 0.1$ & $0.7 \pm 0.3$ & $6.70 \times 10^{10}$ & 4.9 & $0.8234 \pm 0.0000$ \\
74259 & 31.783893 & 15.337084 & $-18.6 \pm 0.2$ & $0.8 \pm 0.2$ & $1.45 \times 10^{10}$ & 2.2 & $0.3465 \pm 0.0003$ \\
11294 & 117.68273 & 14.671243 & $-15.1 \pm 0.1$ & $0.6 \pm 0.2$ & $1.69 \times 10^{8}$ & 1.2 & $0.1084 \pm 0.0003$ \\
96005 & 58.933431 & 9.6326215 & $-17.7 \pm 0.1$ & $0.7 \pm 0.2$ & $2.80 \times 10^{9}$ & 2.8 & $0.2577 \pm 0.0002$ \\
94241 & 58.883431 & 9.7031105 & $-19.9 \pm 0.1$ & $0.5 \pm 0.3$ & $1.36 \times 10^{10}$ & 3.3 & $0.5974 \pm 0.0003$ \\
50723 & 258.05127 & 33.519435 & $-19.9 \pm 0.2$ & $0.5 \pm 0.3$ & $9.97 \times 10^{9}$ & 3.2 & $0.6327 \pm 0.0001$ \\
97319 & 58.937671 & 9.6354616 & $-19.0 \pm 0.2$ & $0.4 \pm 0.3$ & $3.09 \times 10^{9}$ & 5.3 & $0.5565 \pm 0.0004$ \\
70697 & 31.69978 & 15.379112 & $-18.7 \pm 0.1$ & $0.7 \pm 0.1$ & $5.53 \times 10^{9}$ & 3.4 & $0.2588 \pm 0.0003$ \\
63536 & 13.388875 & 12.554246 & $-19.1 \pm 0.1$ & $0.4 \pm 0.2$ & $2.71 \times 10^{9}$ & 4.2 & $0.4324 \pm 0.0002$ \\
25137 & 177.64874 & 28.817094 & $-22.3 \pm 0.1$ & $0.8 \pm 0.3$ & $2.68 \times 10^{11}$ & 3.9 & $0.9552 \pm 0.0007$ \\
34354 & 218.83454 & 24.935074 & $-19.3 \pm 0.2$ & $0.3 \pm 0.4$ & $1.37 \times 10^{9}$ & 4.5 & $0.5764 \pm 0.0002$ \\
62009 & 13.336815 & 12.585924 & $-20.5 \pm 0.1$ & $0.2 \pm 0.2$ & $4.10 \times 10^{9}$ & 10.8 & $0.7693 \pm 0.0003$ \\
34876 & 218.84784 & 24.90464 & $-20.4 \pm 0.2$ & $0.4 \pm 0.3$ & $7.31 \times 10^{9}$ & 4.3 & $0.6930 \pm 0.0004$ \\
24641 & 177.63353 & 28.826078 & $-22.4 \pm 0.1$ & $0.8 \pm 0.2$ & $4.12 \times 10^{11}$ & 7.7 & $0.9255 \pm 0.0007$ \\
73794 & 31.771565 & 15.407626 & $-22.9 \pm 0.1$ & $0.9 \pm 0.2$ & $8.62 \times 10^{11}$ & 5.5 & $1.0372 \pm 0.0016$ \\
50760 & 258.05436 & 33.518538 & $-18.8 \pm 0.2$ & $0.6 \pm 0.3$ & $5.15 \times 10^{9}$ & 4.2 & $0.3734 \pm 0.0002$ \\
72265 & 31.735111 & 15.422063 & $-21.5 \pm 0.1$ & $0.3 \pm 0.2$ & $1.07 \times 10^{10}$ & 4.6 & $0.8162 \pm 0.0003$ \\
97354 & 58.940483 & 9.6255704 & $-17.2 \pm 0.2$ & $0.8 \pm 0.3$ & $3.07 \times 10^{9}$ & 2.1 & $0.2563 \pm 0.0005$ \\
62539 & 13.360299 & 12.59699 & $-19.8 \pm 0.1$ & $0.6 \pm 0.3$ & $1.00 \times 10^{10}$ & 4.8 & $0.6089 \pm 0.0001$ \\
41551 & 58.853293 & 9.7468418 & $-20.7 \pm 0.2$ & $0.9 \pm 0.3$ & $7.83 \times 10^{10}$ & 3.2 & $0.6885 \pm 0.0009$ \\
33872 & 218.82435 & 24.871632 & $-17.7 \pm 0.1$ & $0.4 \pm 0.2$ & $6.90 \times 10^{8}$ & 4.0 & $0.2530 \pm 0.0002$ \\
35409 & 218.86221 & 24.946501 & $-19.9 \pm 0.2$ & $0.2 \pm 0.3$ & $2.11 \times 10^{9}$ & 6.1 & $0.6201 \pm 0.0004$ \\
51795 & 258.15643 & 33.565313 & $-19.1 \pm 0.1$ & $0.5 \pm 0.2$ & $6.13 \times 10^{9}$ & 4.7 & $0.4100 \pm 0.0000$ \\
83546 & 327.72665 & 28.914356 & $-19.8 \pm 0.1$ & $0.7 \pm 0.2$ & $2.71 \times 10^{10}$ & 3.0 & $0.4903 \pm 0.0003$ \\
51295 & 258.10443 & 33.568475 & $-16.6 \pm 0.1$ & $0.4 \pm 0.2$ & $3.55 \times 10^{8}$ & 1.5 & $0.1383 \pm 0.0002$ \\
94820 & 58.899337 & 9.6807377 & $-21.5 \pm 0.2$ & $0.7 \pm 0.3$ & $6.39 \times 10^{10}$ & 3.5 & $0.8359 \pm 0.0001$ \\
24166 & 177.61837 & 28.740747 & $-18.6 \pm 0.2$ & $0.5 \pm 0.4$ & $2.51 \times 10^{9}$ & 3.9 & $0.4461 \pm 0.0002$ \\
25706 & 177.66375 & 28.767888 & $-18.6 \pm 0.2$ & $0.6 \pm 0.3$ & $3.23 \times 10^{9}$ & 3.6 & $0.4167 \pm 0.0002$ \\
62573 & 13.363634 & 12.552785 & $-19.8 \pm 0.1$ & $0.5 \pm 0.1$ & $8.02 \times 10^{9}$ & 2.7 & $0.4404 \pm 0.0002$ \\
83568 & 327.72845 & 28.926887 & $-17.1 \pm 0.1$ & $0.6 \pm 0.2$ & $1.22 \times 10^{9}$ & 2.6 & $0.2110 \pm 0.0003$ \\
10868 & 117.66269 & 14.684116 & $-19.2 \pm 0.1$ & $0.4 \pm 0.2$ & $3.21 \times 10^{9}$ & 3.1 & $0.4416 \pm 0.0001$ \\
45685 & 58.939923 & 9.7243562 & $-20.0 \pm 0.1$ & $0.6 \pm 0.2$ & $2.25 \times 10^{10}$ & 2.6 & $0.5965 \pm 0.0003$ \\
76409 & 31.81002 & 15.350278 & $-17.9 \pm 0.1$ & $0.5 \pm 0.1$ & $2.11 \times 10^{9}$ & 2.1 & $0.2124 \pm 0.0002$ \\
51322 & 258.10705 & 33.555848 & $-20.6 \pm 0.1$ & $0.4 \pm 0.1$ & $8.83 \times 10^{9}$ & 2.8 & $0.4664 \pm 0.0002$ \\
93822 & 58.873397 & 9.685874 & $-20.4 \pm 0.1$ & $0.7 \pm 0.2$ & $5.08 \times 10^{10}$ & 12.2 & $0.4751 \pm 0.0003$ \\
97408 & 58.945616 & 9.6961655 & $-18.2 \pm 0.2$ & $0.7 \pm 0.3$ & $4.29 \times 10^{9}$ & 2.3 & $0.3700 \pm 0.0004$ \\
51849 & 258.1588 & 33.528547 & $-20.3 \pm 0.1$ & $0.6 \pm 0.2$ & $2.68 \times 10^{10}$ & 3.4 & $0.4364 \pm 0.0003$ \\
84110 & 327.72675 & 28.898628 & $-17.6 \pm 0.1$ & $0.7 \pm 0.1$ & $3.42 \times 10^{9}$ & 1.3 & $0.1909 \pm 0.0002$ \\
73359 & 31.76205 & 15.374347 & $-18.8 \pm 0.2$ & $0.3 \pm 0.3$ & $1.11 \times 10^{9}$ & 3.3 & $0.4531 \pm 0.0001$ \\
36497 & 218.86883 & 24.889403 & $-16.8 \pm 0.2$ & $0.4 \pm 0.4$ & $2.68 \times 10^{8}$ & 2.5 & $0.2660 \pm 0.0004$ \\
41667 & 58.856774 & 9.7469731 & $-19.9 \pm 0.1$ & $0.9 \pm 0.3$ & $5.34 \times 10^{10}$ & 3.9 & $0.5823 \pm 0.0008$ \\
72343 & 31.738636 & 15.393939 & $-17.9 \pm 0.1$ & $0.6 \pm 0.1$ & $1.97 \times 10^{9}$ & 2.1 & $0.2012 \pm 0.0002$ \\
74392 & 31.785745 & 15.374634 & $-20.1 \pm 0.1$ & $0.5 \pm 0.2$ & $1.29 \times 10^{10}$ & 2.7 & $0.6998 \pm 0.0003$ \\
25253 & 177.65488 & 28.829554 & $-16.9 \pm 0.1$ & $0.5 \pm 0.2$ & $7.66 \times 10^{8}$ & 5.0 & $0.1917 \pm 0.0002$ \\
32934 & 218.79974 & 24.884314 & $-21.2 \pm 0.2$ & $0.6 \pm 0.4$ & $3.65 \times 10^{10}$ & 7.8 & $0.9080 \pm 0.0000$ \\
74409 & 31.786186 & 15.392032 & $-18.2 \pm 0.2$ & $0.8 \pm 0.3$ & $6.98 \times 10^{9}$ & 2.0 & $0.3637 \pm 0.0002$ \\
51374 & 258.11168 & 33.556927 & $-18.7 \pm 0.2$ & $0.6 \pm 0.4$ & $4.56 \times 10^{9}$ & 2.7 & $0.4087 \pm 0.0003$ \\
94898 & 58.90221 & 9.6712301 & $-15.3 \pm 0.1$ & $0.7 \pm 0.2$ & $3.80 \times 10^{8}$ & 0.9 & $0.0839 \pm 0.0003$ \\
83126 & 327.69152 & 28.899508 & $-19.5 \pm 0.1$ & $0.7 \pm 0.2$ & $2.18 \times 10^{10}$ & 3.5 & $0.3772 \pm 0.0002$ \\
96457 & 58.968276 & 9.6532634 & $-19.2 \pm 0.2$ & $0.6 \pm 0.4$ & $1.20 \times 10^{10}$ & 5.4 & $0.5657 \pm 0.0001$ \\
61624 & 13.318385 & 12.611105 & $-19.5 \pm 0.2$ & $0.6 \pm 0.3$ & $1.19 \times 10^{10}$ & 2.9 & $0.6222 \pm 0.0002$ \\
45242 & 58.951462 & 9.7563091 & $-21.8 \pm 0.1$ & $0.9 \pm 0.2$ & $3.75 \times 10^{11}$ & 5.7 & $0.8689 \pm 0.0004$ \\
61116 & 13.295006 & 12.618442 & $-18.7 \pm 0.2$ & $0.6 \pm 0.3$ & $4.55 \times 10^{9}$ & 4.3 & $0.4106 \pm 0.0001$ \\
24253 & 177.6215 & 28.752339 & $-20.1 \pm 0.2$ & $0.5 \pm 0.3$ & $1.62 \times 10^{10}$ & 3.7 & $0.6177 \pm 0.0005$ \\
83649 & 327.7268 & 28.920093 & $-18.1 \pm 0.1$ & $0.8 \pm 0.1$ & $8.42 \times 10^{9}$ & 2.6 & $0.2105 \pm 0.0003$ \\
94403 & 58.891081 & 9.6869853 & $-18.0 \pm 0.1$ & $0.9 \pm 0.3$ & $1.49 \times 10^{10}$ & 2.1 & $0.3097 \pm 0.0002$ \\
35529 & 218.86359 & 24.944 & $-20.9 \pm 0.2$ & $0.3 \pm 0.3$ & $8.35 \times 10^{9}$ & 5.0 & $0.8679 \pm 0.0000$ \\
12491 & 117.74234 & 14.620434 & $-18.0 \pm 0.1$ & $0.2 \pm 0.3$ & $3.00 \times 10^{8}$ & 2.7 & $0.3317 \pm 0.0002$ \\
11992 & 117.71671 & 14.632345 & $-22.7 \pm 0.1$ & $0.9 \pm 0.2$ & $8.37 \times 10^{11}$ & 7.4 & $1.2066 \pm 0.0011$ \\
10963 & 117.66855 & 14.62808 & $-20.4 \pm 0.1$ & $0.7 \pm 0.2$ & $4.14 \times 10^{10}$ & 5.1 & $0.6294 \pm 0.0002$ \\
73940 & 31.775236 & 15.392164 & $-20.5 \pm 0.2$ & $0.3 \pm 0.3$ & $4.06 \times 10^{9}$ & 2.7 & $0.7116 \pm 0.0002$ \\
96469 & 58.966364 & 9.6745306 & $-17.7 \pm 0.1$ & $0.7 \pm 0.2$ & $3.22 \times 10^{9}$ & 4.3 & $0.2656 \pm 0.0004$ \\
24280 & 177.62182 & 28.761392 & $-20.1 \pm 0.2$ & $0.4 \pm 0.4$ & $6.23 \times 10^{9}$ & 4.8 & $0.7069 \pm 0.0000$ \\
34524 & 218.84182 & 24.862797 & $-19.4 \pm 0.2$ & $0.2 \pm 0.4$ & $1.39 \times 10^{9}$ & 4.6 & $0.6033 \pm 0.0010$ \\
81630 & 327.63202 & 28.862008 & $-19.8 \pm 0.2$ & $0.5 \pm 0.4$ & $7.19 \times 10^{9}$ & 7.7 & $0.6161 \pm 0.0000$ \\
61673 & 13.322306 & 12.583941 & $-20.7 \pm 0.1$ & $0.4 \pm 0.2$ & $9.10 \times 10^{9}$ & 5.5 & $0.6584 \pm 0.0004$ \\
37099 & 218.8767 & 24.955592 & $-18.2 \pm 0.2$ & $0.5 \pm 0.4$ & $1.76 \times 10^{9}$ & 3.0 & $0.3784 \pm 0.0002$ \\
95980 & 58.932353 & 9.6729633 & $-15.0 \pm 0.2$ & $0.6 \pm 0.3$ & $1.59 \times 10^{8}$ & 2.0 & $0.1286 \pm 0.0003$ \\
23790 & 177.60787 & 28.838315 & $-21.6 \pm 0.1$ & $0.4 \pm 0.3$ & $2.79 \times 10^{10}$ & 6.6 & $0.7795 \pm 0.0013$ \\
23281 & 177.59117 & 28.764748 & $-19.7 \pm 0.2$ & $0.3 \pm 0.3$ & $2.46 \times 10^{9}$ & 3.8 & $0.6226 \pm 0.0003$ \\
61175 & 13.298055 & 12.617147 & $-17.3 \pm 0.1$ & $1.0 \pm 0.2$ & $7.36 \times 10^{9}$ & 2.3 & $0.2239 \pm 0.0002$ \\
51962 & 258.15061 & 33.514475 & $-20.0 \pm 0.2$ & $0.5 \pm 0.3$ & $7.37 \times 10^{9}$ & 3.5 & $0.6443 \pm 0.0002$ \\
62715 & 13.369726 & 12.548216 & $-20.7 \pm 0.1$ & $0.6 \pm 0.1$ & $3.68 \times 10^{10}$ & 5.8 & $0.5093 \pm 0.0004$ \\
95485 & 58.919852 & 9.6494973 & $-15.2 \pm 0.1$ & $0.5 \pm 0.2$ & $1.82 \times 10^{8}$ & 2.1 & $0.0839 \pm 0.0002$ \\
42238 & 58.873838 & 9.723309 & $-20.4 \pm 0.1$ & $0.8 \pm 0.2$ & $5.13 \times 10^{10}$ & 6.7 & $0.5815 \pm 0.0005$ \\
62933 & 13.377116 & 12.609484 & $-18.0 \pm 0.1$ & $0.5 \pm 0.2$ & $1.36 \times 10^{9}$ & 1.8 & $0.2789 \pm 0.0001$ \\
72453 & 31.738936 & 15.419616 & $-20.7 \pm 0.1$ & $0.1 \pm 0.2$ & $3.09 \times 10^{9}$ & 5.6 & $0.8154 \pm 0.0004$ \\
74503 & 31.790136 & 15.362447 & $-20.8 \pm 0.1$ & $0.2 \pm 0.2$ & $3.62 \times 10^{9}$ & 5.5 & $0.8770 \pm 0.0000$ \\
10504 & 117.6492 & 14.625036 & $-20.3 \pm 0.1$ & $1.0 \pm 0.2$ & $1.21 \times 10^{11}$ & 4.9 & $0.4430 \pm 0.0011$ \\
35084 & 218.85304 & 24.8682 & $-20.2 \pm 0.1$ & $0.3 \pm 0.2$ & $5.95 \times 10^{9}$ & 4.2 & $0.6023 \pm 0.0002$ \\
12049 & 117.71996 & 14.674784 & $-16.8 \pm 0.1$ & $1.0 \pm 0.2$ & $7.12 \times 10^{9}$ & 2.9 & $0.1609 \pm 0.0000$ \\
82196 & 327.6544 & 28.891059 & $-21.3 \pm 0.1$ & $0.3 \pm 0.2$ & $9.45 \times 10^{9}$ & 5.8 & $0.7882 \pm 0.0007$ \\
62231 & 13.347263 & 12.599827 & $-17.9 \pm 0.0$ & $0.5 \pm 0.1$ & $2.25 \times 10^{9}$ & 1.8 & $0.1814 \pm 0.0002$ \\
35097 & 218.85263 & 24.964962 & $-19.6 \pm 0.2$ & $0.2 \pm 0.4$ & $1.63 \times 10^{9}$ & 5.9 & $0.6626 \pm 0.0002$ \\
44826 & 58.963056 & 9.7607149 & $-21.3 \pm 0.1$ & $0.7 \pm 0.2$ & $8.39 \times 10^{10}$ & 4.8 & $0.7182 \pm 0.0000$ \\
33157 & 218.80589 & 24.921957 & $-19.8 \pm 0.2$ & $0.2 \pm 0.3$ & $1.69 \times 10^{9}$ & 4.6 & $0.6210 \pm 0.0002$ \\
23840 & 177.60944 & 28.80686 & $-20.2 \pm 0.1$ & $0.4 \pm 0.3$ & $8.53 \times 10^{9}$ & 6.3 & $0.6201 \pm 0.0005$ \\
71458 & 31.714661 & 15.397698 & $-19.3 \pm 0.1$ & $0.8 \pm 0.2$ & $2.85 \times 10^{10}$ & 6.1 & $0.3553 \pm 0.0003$ \\
61221 & 13.301121 & 12.589168 & $-18.3 \pm 0.1$ & $0.5 \pm 0.2$ & $1.95 \times 10^{9}$ & 2.8 & $0.2793 \pm 0.0002$ \\
62758 & 13.369905 & 12.578817 & $-20.2 \pm 0.1$ & $0.5 \pm 0.2$ & $1.16 \times 10^{10}$ & 5.8 & $0.6265 \pm 0.0002$ \\
23602 & 177.60137 & 28.78474 & $-18.2 \pm 0.2$ & $0.7 \pm 0.4$ & $5.38 \times 10^{9}$ & 3.3 & $0.3928 \pm 0.0003$ \\
72495 & 31.741848 & 15.373375 & $-20.6 \pm 0.2$ & $0.5 \pm 0.3$ & $1.69 \times 10^{10}$ & 6.1 & $0.6968 \pm 0.0018$ \\
44338 & 58.928732 & 9.7278052 & $-21.1 \pm 0.1$ & $0.8 \pm 0.2$ & $1.37 \times 10^{11}$ & 3.2 & $0.7081 \pm 0.0003$ \\
73011 & 31.754217 & 15.392764 & $-18.9 \pm 0.2$ & $0.8 \pm 0.2$ & $2.03 \times 10^{10}$ & 2.9 & $0.3552 \pm 0.0002$ \\
51208 & 258.09604 & 33.552338 & $-17.6 \pm 0.2$ & $0.3 \pm 0.3$ & $3.86 \times 10^{8}$ & 2.2 & $0.2612 \pm 0.0003$ \\
33592 & 218.81804 & 24.927287 & $-9.4 \pm 0.2$ & $1.0 \pm 0.3$ & $7.20 \times 10^{6}$ & 0.2 & $0.0111 \pm 0.0001$ \\
24889 & 177.64213 & 28.777686 & $-21.1 \pm 0.1$ & $0.8 \pm 0.3$ & $1.11 \times 10^{11}$ & 4.2 & $0.6534 \pm 0.0003$ \\
24890 & 177.64286 & 28.822384 & $-14.3 \pm 0.1$ & $0.3 \pm 0.1$ & $2.59 \times 10^{7}$ & 1.2 & $0.0435 \pm 0.0005$ \\
71483 & 31.715843 & 15.391475 & $-20.5 \pm 0.1$ & $0.3 \pm 0.2$ & $6.18 \times 10^{9}$ & 4.2 & $0.5948 \pm 0.0002$ \\
81728 & 327.63752 & 28.841285 & $-15.0 \pm 0.1$ & $0.7 \pm 0.2$ & $2.93 \times 10^{8}$ & 1.6 & $0.1110 \pm 0.0003$ \\
62274 & 13.349554 & 12.560545 & $-19.6 \pm 0.1$ & $0.8 \pm 0.2$ & $3.76 \times 10^{10}$ & 2.2 & $0.4316 \pm 0.0007$ \\
33093 & 218.80474 & 24.881328 & $-20.6 \pm 0.1$ & $0.5 \pm 0.3$ & $1.77 \times 10^{10}$ & 4.1 & $0.5937 \pm 0.0005$ \\
34122 & 218.8305 & 24.978486 & $-20.3 \pm 0.2$ & $0.7 \pm 0.3$ & $4.86 \times 10^{10}$ & 6.1 & $0.6614 \pm 0.0004$ \\
11083 & 117.67438 & 14.623368 & $-20.6 \pm 0.1$ & $0.7 \pm 0.2$ & $5.19 \times 10^{10}$ & 3.7 & $0.6638 \pm 0.0002$ \\
62285 & 13.347985 & 12.626399 & $-20.5 \pm 0.1$ & $0.2 \pm 0.3$ & $4.13 \times 10^{9}$ & 6.1 & $0.8057 \pm 0.0003$ \\
61263 & 13.301853 & 12.611254 & $-20.3 \pm 0.1$ & $0.4 \pm 0.2$ & $9.72 \times 10^{9}$ & 4.4 & $0.6264 \pm 0.0002$ \\
60752 & 13.281069 & 12.591079 & $-16.8 \pm 0.1$ & $0.4 \pm 0.1$ & $2.83 \times 10^{8}$ & 1.5 & $0.1304 \pm 0.0004$ \\
93746 & 58.869862 & 9.7265934 & $-22.2 \pm 0.1$ & $0.8 \pm 0.3$ & $1.65 \times 10^{11}$ & 19.8 & $0.8942 \pm 0.0001$ \\
97593 & 58.969292 & 9.6737295 & $-19.0 \pm 0.1$ & $0.7 \pm 0.2$ & $9.87 \times 10^{9}$ & 3.2 & $0.3097 \pm 0.0002$ \\
95576 & 58.919818 & 9.669419 & $-17.1 \pm 0.2$ & $0.8 \pm 0.4$ & $2.71 \times 10^{9}$ & 1.9 & $0.2997 \pm 0.0001$ \\
51038 & 258.07973 & 33.554497 & $-18.3 \pm 0.2$ & $0.4 \pm 0.3$ & $1.03 \times 10^{9}$ & 2.8 & $0.3741 \pm 0.0004$ \\
75615 & 31.824744 & 15.381646 & $-17.1 \pm 0.1$ & $0.6 \pm 0.2$ & $6.98 \times 10^{8}$ & 1.4 & $0.2141 \pm 0.0003$ \\
83089 & 327.68992 & 28.896092 & $-19.2 \pm 0.1$ & $0.6 \pm 0.2$ & $7.15 \times 10^{9}$ & 2.3 & $0.3625 \pm 0.0003$ \\
95593 & 58.920976 & 9.679905 & $-20.6 \pm 0.1$ & $0.5 \pm 0.3$ & $2.21 \times 10^{10}$ & 4.0 & $0.7002 \pm 0.0004$ \\
51563 & 258.13133 & 33.570441 & $-18.3 \pm 0.1$ & $0.4 \pm 0.2$ & $1.03 \times 10^{9}$ & 3.8 & $0.2721 \pm 0.0001$ \\
61804 & 13.328091 & 12.580477 & $-17.9 \pm 0.2$ & $0.9 \pm 0.3$ & $9.41 \times 10^{9}$ & 4.6 & $0.3704 \pm 0.0002$ \\
74097 & 31.779667 & 15.367699 & $-21.7 \pm 0.1$ & $0.7 \pm 0.2$ & $9.53 \times 10^{10}$ & 8.0 & $0.9356 \pm 0.0011$ \\
62834 & 13.373869 & 12.541097 & $-20.1 \pm 0.1$ & $0.6 \pm 0.2$ & $1.46 \times 10^{10}$ & 2.7 & $0.5271 \pm 0.0003$ \\
50835 & 258.0621 & 33.531448 & $-18.2 \pm 0.1$ & $0.5 \pm 0.2$ & $1.77 \times 10^{9}$ & 2.7 & $0.2891 \pm 0.0003$ \\
81781 & 327.63775 & 28.890182 & $-20.1 \pm 0.2$ & $0.7 \pm 0.3$ & $2.59 \times 10^{10}$ & 4.6 & $0.6360 \pm 0.0008$ \\
34793 & 218.84623 & 24.908811 & $-19.2 \pm 0.2$ & $0.4 \pm 0.3$ & $4.43 \times 10^{9}$ & 3.8 & $0.5664 \pm 0.0003$ \\
43384 & 58.902215 & 9.7538971 & $-19.4 \pm 0.2$ & $0.7 \pm 0.3$ & $1.50 \times 10^{10}$ & 5.7 & $0.5850 \pm 0.0000$ \\
84348 & 327.69791 & 28.886163 & $-19.2 \pm 0.1$ & $0.7 \pm 0.2$ & $1.15 \times 10^{10}$ & 8.0 & $0.3637 \pm 0.0003$ \\
61822 & 13.327973 & 12.615272 & $-17.1 \pm 0.1$ & $0.4 \pm 0.2$ & $7.53 \times 10^{8}$ & 2.0 & $0.2666 \pm 0.0001$ \\
23425 & 177.59522 & 28.798456 & $-21.1 \pm 0.1$ & $0.4 \pm 0.3$ & $1.21 \times 10^{10}$ & 4.5 & $0.8253 \pm 0.0000$ \\
35203 & 218.85581 & 24.895004 & $-20.1 \pm 0.2$ & $0.2 \pm 0.3$ & $2.64 \times 10^{9}$ & 3.9 & $0.6755 \pm 0.0003$ \\
76677 & 31.806422 & 15.369457 & $-21.2 \pm 0.1$ & $0.3 \pm 0.2$ & $7.87 \times 10^{9}$ & 4.5 & $0.7702 \pm 0.0002$ \\
61837 & 13.329307 & 12.593104 & $-19.3 \pm 0.2$ & $0.4 \pm 0.3$ & $3.18 \times 10^{9}$ & 3.5 & $0.5459 \pm 0.0003$ \\
12174 & 117.72508 & 14.683515 & $-19.2 \pm 0.1$ & $0.5 \pm 0.2$ & $4.94 \times 10^{9}$ & 2.1 & $0.4969 \pm 0.0003$ \\
23446 & 177.59583 & 28.800836 & $-19.9 \pm 0.1$ & $0.5 \pm 0.3$ & $1.14 \times 10^{10}$ & 3.6 & $0.6077 \pm 0.0004$ \\
93083 & 58.851263 & 9.6678248 & $-19.1 \pm 0.1$ & $0.7 \pm 0.2$ & $8.43 \times 10^{9}$ & 3.7 & $0.3431 \pm 0.0001$ \\
51416 & 258.11648 & 33.531677 & $-20.0 \pm 0.1$ & $0.5 \pm 0.2$ & $1.39 \times 10^{10}$ & 4.3 & $0.4643 \pm 0.0004$ \\
24989 & 177.6442 & 28.79539 & $-17.7 \pm 0.2$ & $0.5 \pm 0.3$ & $1.05 \times 10^{9}$ & 2.2 & $0.3262 \pm 0.0004$ \\
10654 & 117.65387 & 14.660226 & $-18.0 \pm 0.1$ & $0.5 \pm 0.2$ & $1.93 \times 10^{9}$ & 2.0 & $0.3026 \pm 0.0002$ \\
51024 & 258.0792 & 33.566421 & $-18.5 \pm 0.1$ & $0.8 \pm 0.2$ & $7.90 \times 10^{9}$ & 2.4 & $0.2637 \pm 0.0003$ \\
61681 & 13.320943 & 12.628812 & $-19.6 \pm 0.2$ & $0.6 \pm 0.3$ & $7.20 \times 10^{9}$ & 4.9 & $0.5817 \pm 0.0003$ \\
82859 & 327.67937 & 28.908205 & $-21.9 \pm 0.1$ & $0.5 \pm 0.2$ & $1.89 \times 10^{10}$ & 4.5 & $0.5372 \pm 0.0004$ \\
50608 & 258.04125 & 33.556564 & $-17.8 \pm 0.2$ & $0.6 \pm 0.3$ & $1.76 \times 10^{9}$ & 3.5 & $0.2636 \pm 0.0002$ \\
50590 & 258.03852 & 33.578987 & $-17.9 \pm 0.2$ & $0.7 \pm 0.2$ & $2.73 \times 10^{9}$ & 2.3 & $0.2638 \pm 0.0001$ \\
73143 & 31.757921 & 15.379198 & $-16.6 \pm 0.2$ & $1.0 \pm 0.3$ & $5.30 \times 10^{9}$ & 1.5 & $0.2068 \pm 0.0004$ \\
71612 & 31.71664 & 15.411962 & $-18.6 \pm 0.1$ & $0.8 \pm 0.2$ & $1.34 \times 10^{10}$ & 2.2 & $0.3754 \pm 0.0003$ \\
50621 & 258.04097 & 33.529415 & $-17.8 \pm 0.3$ & $0.6 \pm 0.5$ & $2.73 \times 10^{9}$ & 1.9 & $0.3728 \pm 0.0001$ \\
63424 & 13.393496 & 12.577992 & $-16.6 \pm 0.1$ & $0.6 \pm 0.2$ & $6.73 \times 10^{8}$ & 1.2 & $0.1939 \pm 0.0001$ \\
72096 & 31.731097 & 15.356583 & $-20.3 \pm 0.1$ & $0.6 \pm 0.3$ & $2.57 \times 10^{10}$ & 2.7 & $0.7008 \pm 0.0002$ \\
23493 & 177.59777 & 28.772034 & $-21.3 \pm 0.1$ & $0.6 \pm 0.3$ & $4.33 \times 10^{10}$ & 3.7 & $0.8201 \pm 0.0002$ \\
93127 & 58.852831 & 9.6540903 & $-18.1 \pm 0.2$ & $0.5 \pm 0.3$ & $1.48 \times 10^{9}$ & 3.4 & $0.3653 \pm 0.0000$ \\
46030 & 58.966713 & 9.7360881 & $-20.0 \pm 0.1$ & $0.7 \pm 0.3$ & $3.37 \times 10^{10}$ & 2.6 & $0.5582 \pm 0.0005$ \\
82383 & 327.66254 & 28.832236 & $-19.4 \pm 0.1$ & $0.6 \pm 0.2$ & $6.66 \times 10^{9}$ & 5.4 & $0.4023 \pm 0.0001$ \\
51664 & 258.1387 & 33.525933 & $-17.0 \pm 0.3$ & $0.8 \pm 0.5$ & $2.18 \times 10^{9}$ & 2.4 & $0.2714 \pm 0.0010$ \\
72149 & 31.731135 & 15.424677 & $-19.7 \pm 0.2$ & $0.5 \pm 0.3$ & $5.16 \times 10^{9}$ & 5.2 & $0.5952 \pm 0.0003$ \\
51676 & 258.13927 & 33.553372 & $-20.0 \pm 0.2$ & $0.5 \pm 0.4$ & $9.34 \times 10^{9}$ & 3.7 & $0.6830 \pm 0.0012$ \\
82398 & 327.66243 & 28.86767 & $-17.8 \pm 0.1$ & $0.6 \pm 0.1$ & $3.09 \times 10^{9}$ & 2.4 & $0.1922 \pm 0.0002$ \\
61520 & 13.313653 & 12.632228 & $-17.2 \pm 0.2$ & $0.6 \pm 0.3$ & $1.66 \times 10^{9}$ & 2.9 & $0.2896 \pm 0.0003$ \\
61922 & 13.332649 & 12.619137 & $-20.3 \pm 0.1$ & $0.3 \pm 0.2$ & $6.00 \times 10^{9}$ & 4.5 & $0.6264 \pm 0.0001$ \\
50939 & 258.0701 & 33.546055 & $-18.0 \pm 0.1$ & $0.6 \pm 0.2$ & $3.07 \times 10^{9}$ & 3.2 & $0.2601 \pm 0.0002$ \\
11236 & 117.6798 & 14.661839 & $-19.1 \pm 0.2$ & $0.5 \pm 0.3$ & $6.54 \times 10^{9}$ & 2.1 & $0.4866 \pm 0.0002$ \\
51177 & 258.09364 & 33.58085 & $-18.1 \pm 0.1$ & $0.5 \pm 0.2$ & $1.51 \times 10^{9}$ & 3.8 & $0.2888 \pm 0.0004$ \\
62968 & 13.379069 & 12.586776 & $-19.1 \pm 0.1$ & $0.5 \pm 0.2$ & $3.35 \times 10^{9}$ & 3.8 & $0.4224 \pm 0.0002$ \\
10748 & 117.65846 & 14.664772 & $-20.1 \pm 0.1$ & $0.8 \pm 0.3$ & $4.70 \times 10^{10}$ & 6.6 & $0.6204 \pm 0.0004$ \\
61950 & 13.335613 & 12.548901 & $-18.5 \pm 0.2$ & $0.6 \pm 0.3$ & $2.69 \times 10^{9}$ & 3.3 & $0.3919 \pm 0.0003$ \\
81919 & 327.64386 & 28.910144 & $-21.6 \pm 0.1$ & $0.4 \pm 0.2$ & $1.70 \times 10^{10}$ & 9.0 & $0.8399 \pm 0.0004$

\enddata

\tablenotetext{a}{WHIQII object number. }
\tablenotetext{b}{Right Ascension (J2000).}
\tablenotetext{c}{Declination (J2000).}
\tablenotetext{d}{K-corrected B-band absolute magnitude.}
\tablenotetext{e}{K-corrected B-V color.}
\tablenotetext{f}{Stellar mass as determined from the mean of the BVRI bands using the \citet{bdj01mstar} M/L relations.}
\tablenotetext{g}{Half-light radius.}
\tablenotetext{h}{Redshift from spectrum.}
\tablecomments{This table is published in its entirety in the electronic edition of the Astrophysical Journal.
A portion is shown here
for guidance regarding itst
form and content.}
\label{tab:photdata}
\end{deluxetable}
\clearpage

\begin{deluxetable}{ccccccc}
\tabletypesize{\scriptsize}

\tablecolumns{7}
\tablecaption{Emission line ratios of WHIQII objects with redshifts.}
\tablehead{
\colhead{WHIQII \#\tablenotemark{a}} &
\colhead{$R_{23}$\tablenotemark{b}} &
\colhead{$O_{32}$\tablenotemark{c}} &
\colhead{$\log(O/H)+12$\tablenotemark{d}} &
\colhead{$R_{23_{\rm EQ}}$\tablenotemark{e}} &
\colhead{$O_{32_{\rm EQ}}$\tablenotemark{f}} &
\colhead{$\log(O/H)_{\rm EQ}+12$\tablenotemark{g}} 
}
\startdata
26114 & \nodata & \nodata & \nodata & \nodata & \nodata & \nodata \\
81835 & \nodata & \nodata & \nodata & \nodata & \nodata & \nodata \\
24584 & \nodata & \nodata & \nodata & \nodata & \nodata & \nodata \\
34826 & \nodata & \nodata & \nodata & \nodata & \nodata & \nodata \\
82434 & $8.3 \pm 0.3$ & $1.2 \pm 0.1$ & $8.5 \pm 0.1$ & $8.5 \pm 0.2$ & $0.9 \pm 0.1$ & $8.5 \pm 0.1$ \\
50703 & $6.9 \pm 0.2$ & $3.8 \pm 0.1$ & $8.7 \pm 0.1$ & $0.0 \pm 0.1$ & $0.3 \pm 0.1$ & \nodata \\
42512 & \nodata & \nodata & \nodata & \nodata & \nodata & \nodata \\
41816 & \nodata & \nodata & \nodata & \nodata & \nodata & \nodata \\
74259 & $3.1 \pm 0.1$ & $1.8 \pm 0.1$ & $9.0 \pm 0.1$ & $4.2 \pm 0.2$ & $0.4 \pm 0.1$ & $8.9 \pm 0.1$ \\
11294 & $10.0 \pm 1.2$ & $1.2 \pm 0.1$ & $8.3 \pm 0.1$ & $12.5 \pm 0.7$ & $0.6 \pm 0.1$ & $8.0 \pm 0.1$ \\
96005 & $10.3 \pm 0.9$ & $5.1 \pm 0.5$ & $8.4 \pm 0.1$ & $0.0 \pm 0.1$ & $0.2 \pm 0.1$ & \nodata \\
94241 & $7.7 \pm 0.2$ & $0.8 \pm 0.1$ & $8.6 \pm 0.1$ & $6.8 \pm 0.1$ & $2.0 \pm 0.3$ & $8.6 \pm 0.1$ \\
50723 & \nodata & \nodata & \nodata & \nodata & \nodata & \nodata \\
97319 & \nodata & \nodata & \nodata & \nodata & \nodata & \nodata \\
70697 & $5.8 \pm 0.3$ & $3.4 \pm 0.1$ & $8.8 \pm 0.1$ & $6.5 \pm 0.1$ & $0.3 \pm 0.1$ & $8.7 \pm 0.1$ \\
63536 & $10.4 \pm 0.5$ & $1.0 \pm 0.1$ & $8.2 \pm 0.1$ & $10.6 \pm 0.1$ & $1.1 \pm 0.1$ & $8.2 \pm 0.1$ \\
25137 & \nodata & \nodata & \nodata & \nodata & \nodata & \nodata \\
34354 & $7.5 \pm 0.2$ & $1.2 \pm 0.1$ & $8.6 \pm 0.1$ & $7.7 \pm 0.2$ & $0.5 \pm 0.1$ & $8.6 \pm 0.1$ \\
62009 & \nodata & \nodata & \nodata & \nodata & \nodata & \nodata \\
34876 & $9.7 \pm 0.4$ & $2.8 \pm 0.1$ & $8.4 \pm 0.1$ & $9.8 \pm 0.1$ & $0.4 \pm 0.1$ & $8.4 \pm 0.1$ \\
24641 & \nodata & \nodata & \nodata & \nodata & \nodata & \nodata \\
73794 & \nodata & \nodata & \nodata & \nodata & \nodata & \nodata \\
50760 & \nodata & \nodata & \nodata & \nodata & \nodata & \nodata \\
72265 & $4.9 \pm 0.1$ & $1.8 \pm 0.1$ & $8.9 \pm 0.1$ & $3.9 \pm 0.2$ & $0.8 \pm 0.1$ & $9.0 \pm 0.1$ \\
97354 & $21.7 \pm 5.5$ & $1.0 \pm 0.1$ & $6.9 \pm 0.6$ & $24.4 \pm 1.4$ & $0.8 \pm 0.1$ & $6.6 \pm 0.1$ \\
62539 & $4.6 \pm 0.3$ & $2.2 \pm 0.1$ & $8.9 \pm 0.1$ & $5.2 \pm 0.3$ & $0.4 \pm 0.1$ & $8.9 \pm 0.1$ \\
41551 & \nodata & \nodata & \nodata & \nodata & \nodata & \nodata \\
33872 & $7.3 \pm 0.1$ & $1.2 \pm 0.1$ & $8.6 \pm 0.1$ & $0.0 \pm 0.1$ & $0.9 \pm 0.1$ & \nodata \\
35409 & \nodata & \nodata & \nodata & \nodata & \nodata & \nodata \\
51795 & \nodata & \nodata & \nodata & \nodata & \nodata & \nodata \\
83546 & $2.3 \pm 0.1$ & $3.2 \pm 0.2$ & $9.0 \pm 0.1$ & $2.5 \pm 0.1$ & $0.3 \pm 0.1$ & $9.0 \pm 0.1$ \\
51295 & $5.8 \pm 0.2$ & $1.1 \pm 0.1$ & $8.2 \pm 0.1$ & $6.1 \pm 0.1$ & $0.9 \pm 0.1$ & $8.7 \pm 0.1$ \\
94820 & \nodata & \nodata & \nodata & \nodata & \nodata & \nodata \\
24166 & $9.2 \pm 0.3$ & $1.0 \pm 0.1$ & $8.4 \pm 0.1$ & $9.7 \pm 0.1$ & $0.9 \pm 0.1$ & $8.3 \pm 0.1$ \\
25706 & $9.3 \pm 0.7$ & $1.6 \pm 0.1$ & $8.4 \pm 0.1$ & $9.0 \pm 0.1$ & $0.7 \pm 0.1$ & $8.4 \pm 0.1$ \\
62573 & $3.5 \pm 0.1$ & $2.4 \pm 0.1$ & $9.0 \pm 0.1$ & $3.8 \pm 0.1$ & $0.4 \pm 0.1$ & $9.0 \pm 0.1$ \\
83568 & $7.4 \pm 0.4$ & $1.8 \pm 0.1$ & $8.6 \pm 0.1$ & $7.6 \pm 0.1$ & $0.6 \pm 0.1$ & $8.6 \pm 0.1$ \\
10868 & $8.8 \pm 0.3$ & $0.5 \pm 0.1$ & $8.4 \pm 0.1$ & $9.8 \pm 0.3$ & $1.9 \pm 0.1$ & $8.3 \pm 0.1$ \\
45685 & $18.5 \pm 1.9$ & $1.1 \pm 0.1$ & $7.3 \pm 0.2$ & $6.8 \pm 0.1$ & $1.6 \pm 0.2$ & $8.7 \pm 5.3$ \\
76409 & $8.6 \pm 0.4$ & $1.8 \pm 0.1$ & $8.5 \pm 0.1$ & $8.4 \pm 0.1$ & $0.6 \pm 0.1$ & $8.5 \pm 0.1$ \\
51322 & $5.4 \pm 0.1$ & $1.6 \pm 0.1$ & $8.8 \pm 0.1$ & $0.0 \pm 0.1$ & $0.7 \pm 0.1$ & \nodata  \\
93822 & $9.9 \pm 0.9$ & $2.9 \pm 0.1$ & $8.4 \pm 0.1$ & $10.9 \pm 0.7$ & $0.3 \pm 0.1$ & $8.3 \pm 0.1$ \\
97408 & $8.9 \pm 0.3$ & $0.6 \pm 0.1$ & $8.4 \pm 0.1$ & $9.0 \pm 0.3$ & $2.2 \pm 0.1$ & $8.4 \pm 0.1$ \\
51849 & $2.0 \pm 0.1$ & $2.7 \pm 0.1$ & $9.0 \pm 0.1$ & $2.4 \pm 0.1$ & $0.3 \pm 0.1$ & $9.0 \pm 0.1$ \\
84110 & $9.1 \pm 0.3$ & $1.2 \pm 0.1$ & $8.4 \pm 0.1$ & $9.0 \pm 0.1$ & $0.9 \pm 0.1$ & $8.4 \pm 0.1$ \\
73359 & $8.3 \pm 0.2$ & $1.6 \pm 0.1$ & $8.5 \pm 0.1$ & $0.0 \pm 0.1$ & $0.5 \pm 0.1$ & \nodata \\
36497 & $17.7 \pm 3.5$ & $0.8 \pm 0.1$ & $7.4 \pm 0.4$ & $18.9 \pm 0.6$ & $1.1 \pm 0.1$ & $7.2 \pm 0.1$ \\
41667 & \nodata & \nodata & \nodata & \nodata & \nodata & \nodata \\
72343 & $6.6 \pm 0.1$ & $1.8 \pm 0.1$ & $8.2 \pm 0.1$ & $0.0 \pm 0.1$ & $0.5 \pm 0.1$ & $-11.8 \pm 0.1$ \\
74392 & $8.4 \pm 0.2$ & $0.6 \pm 0.1$ & $8.5 \pm 0.1$ & $7.7 \pm 0.1$ & $2.8 \pm 0.7$ & $8.5 \pm 0.1$ \\
25253 & $3.9 \pm 0.1$ & $26.3 \pm 0.1$ & $9.0 \pm 0.1$ & $0.0 \pm 0.1$ & $0.0 \pm 0.1$ & $29.5 \pm 4.5$ \\
32934 & \nodata & \nodata & \nodata & \nodata & \nodata & \nodata \\
74409 & \nodata & \nodata & \nodata & \nodata & \nodata & \nodata \\
51374 & $7.8 \pm 0.4$ & $1.3 \pm 0.1$ & $8.6 \pm 0.1$ & $7.6 \pm 0.1$ & $0.9 \pm 0.1$ & $8.6 \pm 0.1$ \\
94898 & $8.9 \pm 0.4$ & $1.8 \pm 0.1$ & $8.5 \pm 0.1$ & $0.0 \pm 0.1$ & $0.5 \pm 0.1$ & \nodata  \\
83126 & $4.6 \pm 0.2$ & $3.1 \pm 0.1$ & $8.9 \pm 0.1$ & $4.8 \pm 0.1$ & $0.3 \pm 0.1$ & $8.9 \pm 0.1$ \\
96457 & \nodata & \nodata & \nodata & \nodata & \nodata & \nodata \\
61624 & $8.5 \pm 0.6$ & $2.3 \pm 0.1$ & $8.5 \pm 0.1$ & $7.9 \pm 0.1$ & $0.5 \pm 0.1$ & $8.6 \pm 0.1$ \\
45242 & \nodata & \nodata & \nodata & \nodata & \nodata & \nodata \\
61116 & $9.6 \pm 0.6$ & $2.0 \pm 0.1$ & $8.4 \pm 0.1$ & $8.8 \pm 0.1$ & $0.6 \pm 0.1$ & $8.5 \pm 0.1$ \\
24253 & $3.3 \pm 0.1$ & $1.2 \pm 0.1$ & $9.0 \pm 0.1$ & $3.8 \pm 0.1$ & $0.8 \pm 0.1$ & $9.0 \pm 0.1$ \\
83649 & $10.4 \pm 1.5$ & $2.8 \pm 0.2$ & $8.3 \pm 0.2$ & $12.9 \pm 0.5$ & $0.3 \pm 0.1$ & $8.0 \pm 0.1$ \\
94403 & $5.2 \pm 0.5$ & $5.1 \pm 0.6$ & $8.9 \pm 0.1$ & $6.4 \pm 0.4$ & $0.2 \pm 0.1$ & $8.8 \pm 0.1$ \\
35529 & \nodata & \nodata & \nodata & \nodata & \nodata & \nodata \\
12491 & $7.0 \pm 0.3$ & $0.3 \pm 0.1$ & $8.6 \pm 0.1$ & $7.4 \pm 0.4$ & $4.1 \pm 0.5$ & $8.5 \pm 0.1$ \\
11992 & \nodata & \nodata & \nodata & \nodata & \nodata & \nodata \\
10963 & $9.1 \pm 0.2$ & $1.1 \pm 0.1$ & $8.4 \pm 0.1$ & $0.0 \pm 0.1$ & $1.6 \pm 0.3$ & \nodata \\
73940 & $6.9 \pm 0.2$ & $1.0 \pm 0.1$ & $8.7 \pm 0.1$ & $7.6 \pm 0.2$ & $1.0 \pm 0.1$ & $8.6 \pm 0.1$ \\
96469 & $6.4 \pm 0.3$ & $1.1 \pm 0.1$ & $8.7 \pm 0.1$ & $5.7 \pm 0.1$ & $1.1 \pm 0.1$ & $8.8 \pm 0.1$ \\
24280 & \nodata & \nodata & \nodata & \nodata & \nodata & \nodata \\
34524 & $3.8 \pm 0.1$ & $2.5 \pm 0.1$ & $9.0 \pm 0.1$ & $0.0 \pm 0.1$ & $0.5 \pm 0.1$ & \nodata \\
81630 & \nodata & \nodata & \nodata & \nodata & \nodata & \nodata \\
61673 & \nodata & \nodata & \nodata & \nodata & \nodata & \nodata \\
37099 & $7.8 \pm 0.6$ & $1.5 \pm 0.1$ & $8.6 \pm 0.1$ & $9.9 \pm 0.8$ & $0.5 \pm 0.1$ & $8.3 \pm 0.1$ \\
95980 & $51.1 \pm 34.2$ & $0.8 \pm 0.1$ & $3.8 \pm 3.3$ & $55.9 \pm 4.2$ & $1.1 \pm 0.2$ & $3.4 \pm 0.4$ \\
23790 & \nodata & \nodata & \nodata & \nodata & \nodata & \nodata \\
23281 & \nodata & \nodata & \nodata & \nodata & \nodata & \nodata \\
61175 & $7.3 \pm 0.1$ & $2.8 \pm 0.2$ & $8.7 \pm 0.1$ & $11.6 \pm 1.1$ & $0.2 \pm 0.1$ & $8.2 \pm 0.1$ \\
51962 & $5.7 \pm 0.3$ & $1.7 \pm 0.1$ & $8.8 \pm 0.1$ & $5.8 \pm 0.1$ & $0.6 \pm 0.1$ & $8.8 \pm 0.1$ \\
62715 & \nodata & \nodata & \nodata & \nodata & \nodata & \nodata \\
95485 & $8.4 \pm 0.8$ & $1.0 \pm 0.1$ & $8.5 \pm 0.1$ & $8.6 \pm 0.2$ & $1.0 \pm 0.1$ & $8.5 \pm 0.1$ \\
42238 & $129.7 \pm 148.6$ & $3.1 \pm 0.1$ & $-1.9 \pm nan$ & $16.8 \pm 2.8$ & $0.4 \pm 0.1$ & $7.6 \pm 0.1$ \\
62933 & $8.6 \pm 0.3$ & $1.4 \pm 0.1$ & $8.4 \pm 0.1$ & $7.9 \pm 0.1$ & $0.9 \pm 0.1$ & $8.5 \pm 0.1$ \\
72453 & $20.0 \pm 1.5$ & $1.0 \pm 0.1$ & $7.1 \pm 0.2$ & $19.0 \pm 0.9$ & $2.0 \pm 0.7$ & $7.2 \pm 0.1$ \\
74503 & \nodata & \nodata & \nodata & \nodata & \nodata & \nodata \\
10504 & \nodata & \nodata & \nodata & \nodata & \nodata & \nodata \\
35084 & $8.1 \pm 0.2$ & $0.5 \pm 0.1$ & $8.5 \pm 0.1$ & $7.6 \pm 1.6$ & $4.8 \pm 3.5$ & $8.5 \pm 0.2$ \\
12049 & \nodata & \nodata & \nodata & \nodata & \nodata & \nodata \\
82196 & \nodata & \nodata & \nodata & \nodata & \nodata & \nodata \\
62231 & $6.1 \pm 0.1$ & $0.8 \pm 0.1$ & $8.7 \pm 0.1$ & $5.4 \pm 0.2$ & $1.9 \pm 0.3$ & $8.8 \pm 0.1$ \\
35097 & $6.3 \pm 0.3$ & $2.5 \pm 0.1$ & $8.8 \pm 0.1$ & $0.0 \pm 0.1$ & $0.4 \pm 0.1$ & \nodata \\
44826 & \nodata & \nodata & \nodata & \nodata & \nodata & \nodata \\
33157 & \nodata & \nodata & \nodata & \nodata & \nodata & \nodata \\
23840 & \nodata & \nodata & \nodata & \nodata & \nodata & \nodata \\
71458 & $5.9 \pm 0.3$ & $2.7 \pm 0.1$ & $8.8 \pm 0.1$ & $0.0 \pm 0.1$ & $0.3 \pm 0.1$ & \nodata \\
61221 & $8.4 \pm 0.4$ & $1.9 \pm 0.1$ & $8.5 \pm 0.1$ & $8.5 \pm 0.1$ & $0.5 \pm 0.1$ & $8.5 \pm 0.1$ \\
62758 & $9.7 \pm 1.1$ & $1.5 \pm 0.1$ & $8.4 \pm 0.1$ & $10.0 \pm 0.2$ & $0.8 \pm 0.1$ & $8.3 \pm 0.1$ \\
23602 & $7.9 \pm 0.6$ & $1.5 \pm 0.1$ & $8.6 \pm 0.1$ & $8.9 \pm 0.3$ & $0.6 \pm 0.1$ & $8.5 \pm 0.1$ \\
72495 & \nodata & \nodata & \nodata & \nodata & \nodata & \nodata \\
44338 & \nodata & \nodata & \nodata & \nodata & \nodata & \nodata \\
73011 & $2.0 \pm 0.1$ & $3.4 \pm 0.2$ & $9.0 \pm 0.1$ & $0.0 \pm 0.1$ & $0.2 \pm 0.1$ & \nodata\\
51208 & $8.9 \pm 0.4$ & $0.8 \pm 0.1$ & $8.4 \pm 0.1$ & $8.7 \pm 0.1$ & $1.4 \pm 0.1$ & $8.4 \pm 0.1$ \\
33592 & \nodata & \nodata & \nodata & \nodata & \nodata & \nodata \\
24889 & $7.4 \pm 0.3$ & $0.9 \pm 0.1$ & $8.6 \pm 0.1$ & $7.9 \pm 0.1$ & $1.0 \pm 0.1$ & $8.5 \pm 0.1$ \\
24890 & $7.8 \pm 0.3$ & $0.9 \pm 0.1$ & $8.6 \pm 0.1$ & $7.5 \pm 0.1$ & $1.3 \pm 0.1$ & $8.6 \pm 0.1$ \\
71483 & $8.2 \pm 0.3$ & $2.5 \pm 0.1$ & $8.6 \pm 0.1$ & $7.8 \pm 0.1$ & $0.5 \pm 0.1$ & $8.6 \pm 0.1$ \\
81728 & $6.5 \pm 0.5$ & $1.9 \pm 0.1$ & $8.7 \pm 0.1$ & $7.2 \pm 0.3$ & $0.5 \pm 0.1$ & $8.7 \pm 0.1$ \\
62274 & $6.9 \pm 0.3$ & $2.7 \pm 0.1$ & $8.7 \pm 0.1$ & $0.0 \pm 0.1$ & $0.2 \pm 0.1$ & \nodata \\
33093 & \nodata & \nodata & \nodata & \nodata & \nodata & \nodata \\
34122 & \nodata & \nodata & \nodata & \nodata & \nodata & \nodata \\
11083 & $8.7 \pm 0.6$ & $1.0 \pm 0.1$ & $8.5 \pm 0.1$ & $8.7 \pm 0.3$ & $1.2 \pm 0.2$ & $8.4 \pm 0.1$ \\
62285 & $3.9 \pm 0.3$ & $2.0 \pm 0.1$ & $9.0 \pm 0.1$ & $3.6 \pm 0.1$ & $0.5 \pm 0.1$ & $9.0 \pm 0.1$ \\
61263 & $7.1 \pm 0.2$ & $1.2 \pm 0.1$ & $8.6 \pm 0.1$ & $7.0 \pm 0.1$ & $1.1 \pm 0.1$ & $8.6 \pm 0.1$ \\
60752 & $8.8 \pm 0.3$ & $0.4 \pm 0.1$ & $8.6 \pm 0.1$ & $9.2 \pm 0.1$ & $3.1 \pm 0.1$ & $8.3 \pm 0.1$ \\
93746 & \nodata & \nodata & \nodata & \nodata & \nodata & \nodata \\
97593 & $9.5 \pm 0.5$ & $1.5 \pm 0.1$ & $8.4 \pm 0.1$ & $10.0 \pm 0.3$ & $0.7 \pm 0.1$ & $8.3 \pm 0.1$ \\
95576 & $9.5 \pm 0.7$ & $1.2 \pm 0.1$ & $8.4 \pm 0.1$ & $10.9 \pm 1.1$ & $0.7 \pm 0.1$ & $8.2 \pm 0.1$ \\
51038 & $6.4 \pm 0.5$ & $0.5 \pm 0.1$ & $8.7 \pm 0.1$ & $6.8 \pm 0.3$ & $1.8 \pm 0.1$ & $8.6 \pm 0.1$ \\
75615 & $12.0 \pm 0.3$ & $1.2 \pm 0.1$ & $8.1 \pm 0.1$ & $0.0 \pm 0.1$ & $1.1 \pm 0.1$ & \nodata \\
83089 & $10.1 \pm 0.4$ & $1.1 \pm 0.1$ & $8.3 \pm 0.1$ & $10.5 \pm 0.1$ & $1.0 \pm 0.1$ & $8.2 \pm 0.1$ \\
95593 & $13.2 \pm 0.3$ & $1.4 \pm 0.1$ & $7.9 \pm 0.1$ & $4.2 \pm 1.4$ & $1.0 \pm 0.1$ & $8.9 \pm 0.1$ \\
51563 & $15.6 \pm 3.6$ & $1.4 \pm 0.1$ & $7.7 \pm 0.4$ & $13.2 \pm 0.4$ & $1.2 \pm 0.2$ & $7.9 \pm 0.1$ \\
61804 & $9.8 \pm 0.7$ & $1.2 \pm 0.1$ & $8.3 \pm 0.1$ & $9.3 \pm 0.1$ & $1.0 \pm 0.1$ & $8.4 \pm 0.1$ \\
74097 & \nodata & \nodata & \nodata & \nodata & \nodata & \nodata \\
62834 & \nodata & \nodata & \nodata & \nodata & \nodata & \nodata \\
50835 & $9.7 \pm 2.3$ & $1.8 \pm 0.1$ & $8.4 \pm 0.3$ & $12.9 \pm 7.3$ & $0.4 \pm 0.2$ & $8.0 \pm 0.6$ \\
81781 & \nodata & \nodata & \nodata & \nodata & \nodata & \nodata \\
34793 & $27.8 \pm 0.4$ & $1.0 \pm 0.1$ & $6.2 \pm 0.1$ & $0.0 \pm 0.1$ & $1.2 \pm 0.1$ & \nodata \\
43384 & \nodata & \nodata & \nodata & \nodata & \nodata & \nodata \\
84348 & $6.2 \pm 0.6$ & $4.2 \pm 0.3$ & $8.8 \pm 0.1$ & $6.9 \pm 0.2$ & $0.2 \pm 0.1$ & $8.7 \pm 0.1$ \\
61822 & $5.2 \pm 0.1$ & $0.1 \pm 0.1$ & $8.8 \pm 0.1$ & $5.3 \pm 0.5$ & $13.5 \pm 9.1$ & $8.8 \pm 0.1$ \\
23425 & \nodata & \nodata & \nodata & \nodata & \nodata & \nodata \\
35203 & $7.5 \pm 0.3$ & $1.4 \pm 0.1$ & $8.6 \pm 0.1$ & $7.0 \pm 0.1$ & $0.9 \pm 0.1$ & $8.6 \pm 0.1$ \\
76677 & \nodata & \nodata & \nodata & \nodata & \nodata & \nodata \\
61837 & $4.3 \pm 0.5$ & $1.9 \pm 0.1$ & $8.9 \pm 0.1$ & $4.1 \pm 0.1$ & $0.6 \pm 0.1$ & $8.9 \pm 0.1$ \\
12174 & $9.0 \pm 0.2$ & $0.4 \pm 0.1$ & $8.4 \pm 0.1$ & $9.5 \pm 0.6$ & $3.9 \pm 0.9$ & $8.3 \pm 0.1$ \\
23446 & \nodata & \nodata & \nodata & \nodata & \nodata & \nodata \\
93083 & $6.3 \pm 0.3$ & $1.8 \pm 0.1$ & $8.7 \pm 0.1$ & $5.7 \pm 0.1$ & $0.7 \pm 0.1$ & $8.8 \pm 0.1$ \\
51416 & $5.6 \pm 0.4$ & $2.8 \pm 0.1$ & $8.8 \pm 0.1$ & $5.8 \pm 0.1$ & $0.4 \pm 0.1$ & $8.8 \pm 0.1$ \\
24989 & $8.0 \pm 0.4$ & $1.3 \pm 0.1$ & $8.5 \pm 0.1$ & $7.4 \pm 0.1$ & $1.0 \pm 0.1$ & $8.6 \pm 0.1$ \\
10654 & $8.5 \pm 0.4$ & $1.0 \pm 0.1$ & $8.5 \pm 0.1$ & $10.1 \pm 0.6$ & $0.8 \pm 0.1$ & $8.3 \pm 0.1$ \\
51024 & \nodata & \nodata & \nodata & \nodata & \nodata & \nodata \\
61681 & $6.7 \pm 0.4$ & $3.1 \pm 0.1$ & $8.7 \pm 0.1$ & $7.5 \pm 0.3$ & $0.3 \pm 0.1$ & $8.6 \pm 0.1$ \\
82859 & $9.2 \pm 0.4$ & $1.5 \pm 0.1$ & $8.4 \pm 0.1$ & $6.0 \pm 1.6$ & $0.4 \pm 0.4$ & $8.8 \pm 0.2$ \\
50608 & \nodata & \nodata & \nodata & \nodata & \nodata & \nodata \\
50590 & \nodata & \nodata & \nodata & \nodata & \nodata & \nodata \\
73143 & $4.3 \pm 0.3$ & $3.6 \pm 0.2$ & $8.9 \pm 0.1$ & $6.8 \pm 0.7$ & $0.2 \pm 0.1$ & $8.7 \pm 0.1$ \\
71612 & $7.7 \pm 0.3$ & $1.7 \pm 0.1$ & $8.6 \pm 0.1$ & $7.8 \pm 0.1$ & $0.6 \pm 0.1$ & $8.6 \pm 0.1$ \\
50621 & \nodata & \nodata & \nodata & \nodata & \nodata & \nodata \\
63424 & $12.9 \pm 0.8$ & $1.6 \pm 0.1$ & $8.0 \pm 0.1$ & $2.4 \pm 0.1$ & $0.0 \pm 0.1$ & $9.0 \pm 0.1$ \\
72096 & \nodata & \nodata & \nodata & \nodata & \nodata & \nodata \\
23493 & \nodata & \nodata & \nodata & \nodata & \nodata & \nodata \\
93127 & $8.9 \pm 0.5$ & $1.2 \pm 0.1$ & $8.4 \pm 0.1$ & $8.1 \pm 0.1$ & $1.2 \pm 0.1$ & $8.5 \pm 0.1$ \\
46030 & $6.9 \pm 0.2$ & $1.1 \pm 0.1$ & $8.7 \pm 0.1$ & $4.9 \pm 1.3$ & $1.7 \pm 0.5$ & $8.9 \pm 0.1$ \\
82383 & $7.0 \pm 0.7$ & $3.1 \pm 0.2$ & $8.7 \pm 0.1$ & $7.5 \pm 0.2$ & $0.3 \pm 0.1$ & $8.6 \pm 0.1$ \\
51664 & \nodata & \nodata & \nodata & \nodata & \nodata & \nodata \\
72149 & \nodata & \nodata & \nodata & \nodata & \nodata & \nodata \\
51676 & \nodata & \nodata & \nodata & \nodata & \nodata & \nodata \\
82398 & $9.6 \pm 0.4$ & $1.2 \pm 0.1$ & $8.4 \pm 0.1$ & $10.0 \pm 0.1$ & $0.8 \pm 0.1$ & $8.3 \pm 0.1$ \\
61520 & $15.0 \pm 3.3$ & $1.4 \pm 0.1$ & $7.7 \pm 0.4$ & $15.1 \pm 0.2$ & $0.8 \pm 0.1$ & $7.7 \pm 0.1$ \\
61922 & $7.5 \pm 0.2$ & $0.9 \pm 0.1$ & $8.6 \pm 0.1$ & $7.3 \pm 0.1$ & $1.3 \pm 0.1$ & $8.6 \pm 0.1$ \\
50939 & \nodata & \nodata & \nodata & \nodata & \nodata & \nodata \\
11236 & $6.5 \pm 0.1$ & $0.5 \pm 0.1$ & $8.7 \pm 0.1$ & $8.3 \pm 1.3$ & $1.4 \pm 0.6$ & $8.5 \pm 0.1$ \\
51177 & $14.9 \pm 1.7$ & $2.1 \pm 0.1$ & $7.8 \pm 0.2$ & $16.4 \pm 0.5$ & $0.4 \pm 0.1$ & $7.6 \pm 0.1$ \\
62968 & $13.5 \pm 1.1$ & $1.5 \pm 0.1$ & $7.9 \pm 0.1$ & $13.8 \pm 0.2$ & $0.7 \pm 0.1$ & $7.9 \pm 0.1$ \\
10748 & $5.8 \pm 0.5$ & $1.3 \pm 0.1$ & $8.8 \pm 0.1$ & $7.1 \pm 0.1$ & $0.6 \pm 0.2$ & $8.7 \pm 0.1$ \\
61950 & \nodata & \nodata & \nodata & \nodata & \nodata & \nodata \\
81919 & \nodata & \nodata & \nodata & \nodata & \nodata & \nodata 

\enddata

\tablenotetext{a}{WHIQII object number. }
\tablenotetext{b}{$R_{23} = \oii+\oiii)/ H\beta$ line flux ratio.}
\tablenotetext{c}{$O_{32} = \oiii / \oii$ line flux ratio.}
\tablenotetext{d}{Oxygen abundance derived from the \citet{KD02} calibration.}
\tablenotetext{e}{$R_{23} = \oii+\oiii)/ H\beta$ equivalent width ratio.}
\tablenotetext{f}{$O_{32} = \oiii / \oii$ equivalent width ratio.  In a few cases this column is missing where continuum measurements were problematic.}
\tablenotetext{g}{Oxygen abundance derived from the \citet{KD02} calibration with equivalent width ratios.}
\tablecomments{This table is published in its entirety in the electronic edition of the Astrophysical Journal.
A portion is shown here
for guidance regarding its
form and content.}
\label{tab:specdata}
\end{deluxetable}

\end{document}